\documentclass[%
 reprint,
 amsmath,amssymb,
 aps,
prb,
]{revtex4-1}

\usepackage{graphicx}
\usepackage{dcolumn}
\usepackage{bm}
\usepackage{version}

\newcommand{\be}{\begin{equation}}
\newcommand{\ee}{\end{equation}}
\newcommand{\bea}{\begin{eqnarray}}
\newcommand{\eea}{\end{eqnarray}}
\newcommand{\Eq}[1]{Eq.\,(\ref{#1})}
\newcommand{\Fig}[1]{Fig.\,\ref{#1}}
\newcommand{\Sec}[1]{Sec.\,\ref{#1}}
\newcommand{\Onlinecite}[1]{Ref.\,[\onlinecite{#1}]} 

\newcommand{\XA}{X_{\rm A}}
\newcommand{\XAT}{X_{\rm A}^-}
\newcommand{\XB}{X_{\rm B}}
\newcommand{\ATPA}{A_{\rm nr}}
\newcommand{\varphiTPA}{\varphi_{\rm nr}}
\newcommand{\TREP}{T_{\rm r}}
\newcommand{\kB}{k_{\rm B}}
\newcommand{\EA}{E_{\rm A}} 

\newcommand{\bK}{\mathbf{K}_{||}}
\newcommand{\MoSe}{SL-MoSe$_2$} 
\newcommand{\taur}{\tau_{\rm r}}

\newcommand{\ie}{i_{\rm e}} 
\newcommand{\ih}{i_{\rm h}} 
\newcommand{\se}{s_{\rm e}} 
\newcommand{\sh}{s_{\rm h}} 

\newcommand{\gv}{\gamma_{\rm v}} 
\newcommand{\gs}{\gamma_{\rm s}} 
\newcommand{\gvs}{\gamma_{\rm vs}} 
\newcommand{\gLT}{\gamma_{\rm LT}} 

\newcommand{\gEEA}{\gamma_{\rm A}} 

\newcommand{\Dc}{\Delta_{\rm cb}}

\begin{document}


\title{Resonantly excited exciton dynamics in two-dimensional $\text{MoSe}_2$ monolayers}
\author{L.\,Scarpelli$^1$}
\author{F.\,Masia$^1$}
\author{E.M.\,Alexeev$^2$}
\author{F.\,Withers$^3$}
\author{A.I.\,Tartakovskii$^2$}
\author{K.S.\,Novoselov$^4$}
\author{W.\,Langbein$^1$}%
\affiliation{
	$^1$School\,of\,Physics\,and\,Astronomy,\,Cardiff\,University,\,The Parade,\,Cardiff\,CF24\,3AA,\,United\,Kingdom\\
	$^2$Department\,of\,Physics\,and\,Astronomy,\,University\,of\,Sheffield,\,Hounsfield\,Rd,\,Sheffield\,S3\,7RH,\,United\,Kingdom\\
	$^3$College\,of\,Engineering,\,Mathematics\,and\,Physical\,Sciences,\,University\,of\,Exeter,\,North\,Park\,Rd,\,Exeter\,EX4\,4QF\,United\,Kingdom\\
    $^4$School\,of\,Physics\,and\,Astronomy,\,University\,of\,Manchester,\,Oxford Rd,\,Manchester\,M13\,9PL,\,United\,Kingdom}%

\date{\today}

\begin{abstract}
\noindent We report on the exciton and trion density dynamics in a single layer of $\text{MoSe}_2$, resonantly excited and probed using three-pulse four-wave mixing (FWM), at temperatures from 300\,K to 77\,K . A multi-exponential third-order response function for amplitude and phase of the heterodyne-detected FWM signal including four decay processes is used to model the data. We provide a consistent interpretation within the intrinsic band structure, not requiring the inclusion of extrinsic effects. We find an exciton radiative lifetime in the sub-picosecond range consistent to what has been recently reported\,\cite{JakubczykNL16}. After the dominating radiative decay, the remaining exciton density, which has been scattered from the initially excited bright radiative state into dark states of different nature by exciton-phonon scattering or disorder scattering, shows a slower dynamics, covering 10ps to 10ns timescales. This includes direct bright transitions with larger in-plane momentum, as well as indirect dark transitions to indirect dark states. We find that exciton-exciton annihilation is not relevant in the observed dynamics, in variance from previous finding under non-resonant excitation. The trion density at 77\,K reveals a decay of the order of 1\,ps, similar to what is observed for the exciton. After few tens of picoseconds, the trion dynamics resembles the one of the exciton, indicating that trion ionization occurs on this timescale. 
\end{abstract}

\maketitle


\section{\label{intro}Introduction}
Single layers (SL) of transition metal dichalcogenides (TMD) can be direct bandgap semiconductors, with the bandgap at the valleys located in the corners K of the hexagonal Brillouin zone\,\cite{MakPRL10,SplendianiNL10}. The spin-orbit interaction and lacking inversion symmetry result in a spin splitting of the electron and hole bands at the two non-equivalent valleys $\pm$K, linking the circular polarization of the transition to the valley index $\pm$\,\cite{XiaoPRL12,MakNNa12}. Therefore, excitons are generated in the +K (-K) valley by absorption of right (left)-handed light, respectively. Due to the reduced screening of the Coulomb interaction and the 2D confinement, excitons in SL-TMD have a large binding energy of a few hundred meV\,\cite{UgedaNMa14}. For the same reason, exciton-exciton interaction is expected to be enhanced, and the effect of exciton-exciton annihilation (EEA) has been reported\,\cite{MouriPRB14,SunNL14,KumarPRB14}. The spin-orbit interaction splits the spin states at the valleys by few hundreds of meV\,\cite{ZhuPRB11} in the valence band (VB), and by few tens of meV\,\cite{KosmiderPRB13} in the conduction band (CB). The exciton transition between the lower CB and higher VB valley is called A exciton ($\XA$) while the transition between the higher CB and lower VB is called B exciton ($\XB$), and has the opposite circular polarization. Mixed transitions are spin forbidden, and transition between different valleys are momentum forbidden and thus called ``dark''. The exciton dynamics therefore depends strongly on the distribution and the scattering between the different dark and bright states and their dispersion\,\cite{ZhangPRL15}. The energy of such dark states depends on the TMD under consideration. In $\rm WX_2$, where $\rm X=S,Se$, optically dark states appear to have a lower energy then bright states, while the opposite is suggested by experiments on $\rm MoX_2$\,\cite{LiuPRB13,KosmiderPRB13}. Dark states are thought to be responsible\,\cite{DeryPRB15} for the anomalous polarization response observed in \MoSe\,\cite{ZhangARX15,DufferwielARX16}. The role of dark states in the exciton dynamics is an important open issue in the understanding of the carrier dynamics. Indeed,  recent experiments on both $\text{MoSe}_2$ and $\text{WSe}_2$ have considered intrinsic dark states to describe the exciton dynamics\,\cite{RobertARX16,SeligNC16}. Population of dark states can occur by exciton-phonon as well as exciton-exciton scattering through many different mechanisms\,\cite{MoodyJOSAB16}. Electron-hole exchange interaction further affects the dynamics by introducing additional energy differences between spin 1 (spin forbidden) and spin 0 (spin-allowed) exciton states\,\cite{YuNC14}.
In the present work we report on resonantly excited exciton and trion density dynamics in a single layer of $\text{MoSe}_2$. The paper is organized as follows: in \Sec{sec:sample} details on the experimental technique and on the investigated sample are given. In \Sec{sec:dephasing} measurements of dephasing time of $\XA$ and the related trion transition are shown for different temperatures. In \Sec{sec:exciton} density dynamics probed at the $\XA$ transition as function of density (\Sec{subsec:excitondensity}) and temperature (\Sec{subsec:excitontemp}) are shown and analyzed, and in \Sec{subsec:excdiscussion} an interpretation of the observed dynamics in terms of the intrinsic band structure is put forward. In \Sec{EEA} the density dependent dynamics  is analyzed for exciton Auger recombination, and no evidence for EEA is found. Finally, in \Sec{Trion dynamics}, we discuss experiments resonant to the trion and in \Sec{Conclusions} we summarize our results.

\section{\label{sec:sample} Sample and experimental setup}

\begin{figure}[bt]
 	\includegraphics[width=\columnwidth]{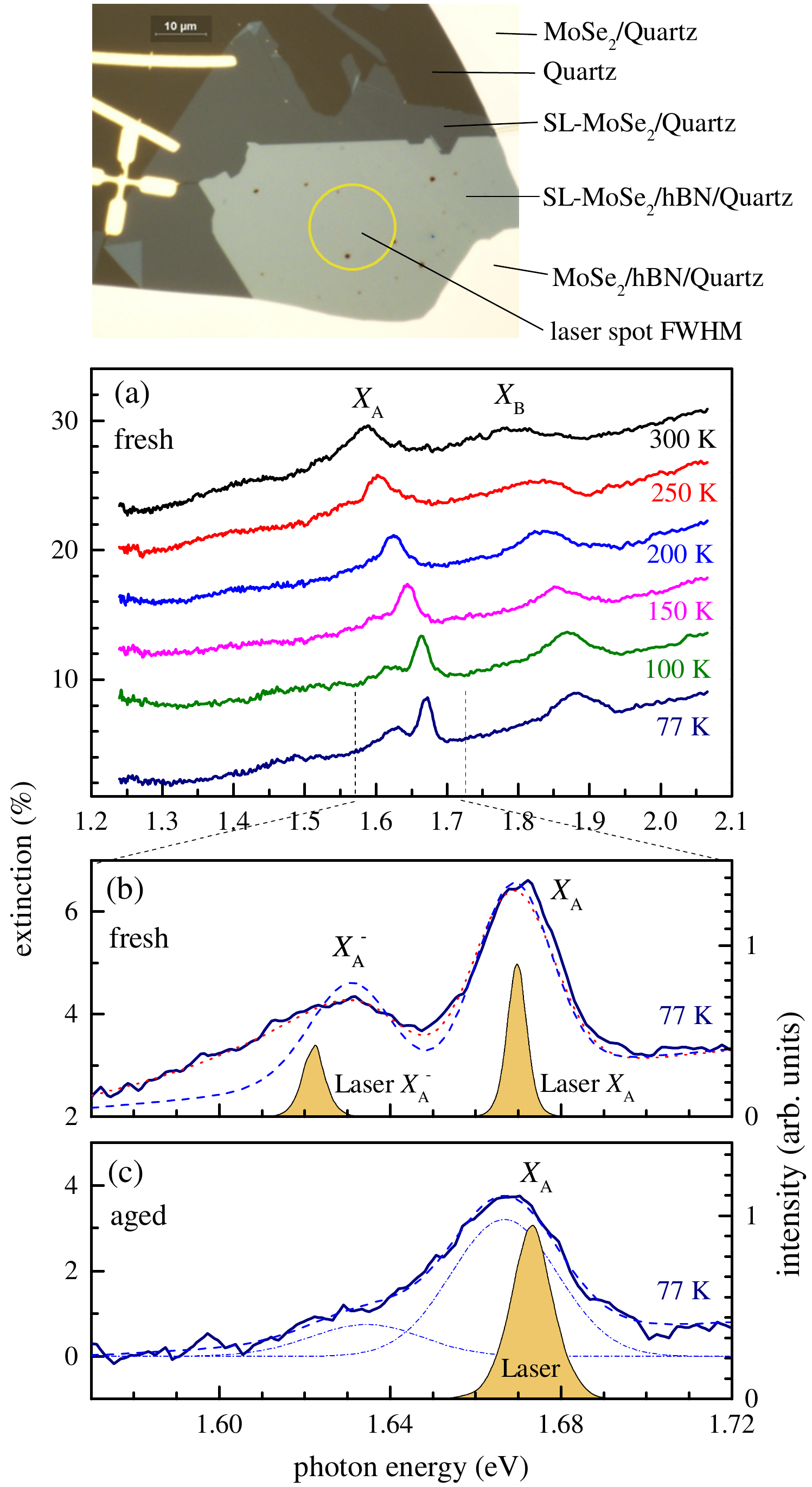}
 	\caption{Sample characterization and excitation laser spectra. Top: Reflection image of the investigated sample, taken with a 0.8 numerical aperture. The different regions are labelled. (a) extinction of the fresh sample as function of temperature, vertically offset for clarity. (b) Zoom of extinction at $T=77$\,K, together with the excitation laser spectrum. (c) as (b), but for the aged sample. Dashed lines are fits of the data (see text). }
 	\label{fig:extinction}
\end{figure}

The sample analyzed in this work is a \MoSe\ transferred onto a  hexagonal BN layer on a 1\,mm thick c-cut quartz substrate. A reflection image is given in the top of \Fig{fig:extinction}. The investigated SL has a size of about $50\times70\,\mu$m$^2$.  The sample is covered by a second 1\,mm thick c-cut quartz substrate to provide thermal radiation screening and mechanical protection, and mounted in a flow cryostat, onto the cold finger in vacuum. 
The extinction of the sample was measured using white-light transmission, referenced to the quartz region adjacent to the investigated flake. The sample was stored in vacuum, and measured 6 months after preparation (``fresh''), and 3 months afterwards, after a few thermal cycles (``aged''). The extinction of the fresh sample as function of temperature during the first cooldown is given in \Fig{fig:extinction}(a). The spectra show the $\XA$ and $\XB$ absorption peaks, which are shifting to higher energies with decreasing temperature as expected. At $T=77\,$K, a peak below the $\XA$ is observed which is attributed to the $\XA$ trion absorption. We call the trion here $\XAT$, assuming a negative charge. Negatively and positively charged trions have very similar binding energies\,\cite{RossNC13}. The trion absorption forms due to the presence of electrons at the edge of the conduction band. The trion absorption observed here shows a low energy tail (see also \Fig{fig:extinction}(b)). In principle a low energy tail is expected for the transition from the thermalized free carrier distribution\,\cite{EsserPRB00, RossNC13}. The trion lineshape in absorption has recently been studied\,\cite{SidlerNPhy16} in a charge-tunable structure  at a temperature of $T=4$\,K, at which the thermal energy of $\kB T\approx 0.34$\,meV is much smaller than the line broadening and such a tail is not observed as expected. At the higher temperature of $T=77\,K$ used here, $\kB T\approx 6.6$\,meV and thus could be relevant. We have fitted the absorption lineshape $A(\hbar\omega)$ using the function
\begin{align} A(x)=&C_{0}+C_{1}x+A_X\exp\left[-\left(\frac{x-E_X}{\sqrt{2}\sigma}\right)^2\right]  \nonumber\\& +\frac{A_T}{2}\left[1-{\rm erf}\left(\frac{x-E_T}{\sqrt{2}\sigma}+\frac{C\sigma}{\sqrt{2}\kB T}\right)\right]\nonumber\\ & \times \exp\left[C\frac{x-E_T}{\kB T}+\left(\frac{C\sigma}{\sqrt{2}\kB T}\right)^2\right]\,.  
\label{eqn:extinction}
\end{align}
$C_0$ and $C_1$ provide a baseline, $A_X$ is the amplitude of the Gaussian exciton peak of position $E_X$, $A_T$ is the amplitude of the trion lineshape of position $E_T$, and $C=M_T^{\ast}/M^{\ast}$ is the ratio between the trion effective mass $M_T^{\ast}=2m_{\rm e}^{\ast}+m_{\rm h}^{\ast}$ and the exciton effective mass $M^{\ast}=m_{\rm e}^{\ast}+m_{\rm h}^{\ast}$. Using the electron effective mass\,\cite{HorzumPRB13} $m_{\rm e}^{\ast}=0.53\,m_0$ and the hole effective mass $m_{\rm h}^{\ast}=0.65\,m_0$, we find $C \approx 1.45$. The width $\sigma$ is the inhomogeneous broadening of exciton and trion, and the trion lineshape is convoluted with an exponential decay due to the thermal electron energy distribution and momentum conservation during absorption. For simplicity we have neglected the electron-energy dependence of the matrix element\,\cite{RossNC13}. A fit to the extinction of the fresh sample (see blue dashed line \Fig{fig:extinction}(b)) shows that the resulting low energy tail of the trion is less extended than observed. We find a trion binding energy of $E_X-E_T\approx 34$\,meV and a FWHM of the exciton peak of $2.355\sigma\approx 20$\,meV. Leaving $C$ as free parameter (see red dotted curve), a good fit is obtained, yielding $C=0.18$, a trion binding energy of 27\,meV and a FWHM of 22\,meV.

A fit to the extinction of the aged sample (see blue dashed line in \Fig{fig:extinction}(c)) with $C=1.45$ shows good agreement with the data, providing a trion binding energy of 28\,meV and a FWHM of 29\,meV. The retrieved trion binding energies are consistent with previous reports \,\cite{RossNC13,SinghPRL14,WangAPL15,JakubczykNL16,GaoPRB16}. The aged sample shows a larger linewidth and a smaller trion contribution. This is about $25\%$ of the exciton peak area, against $63\%$ for the fresh sample using the fit with $C=1.45$. The microscopic origin of the change is unknown - one could speculate that it is due to a wrinkling of the sample after thermal cycling, which creates a larger strain inhomogeneity and different dielectric environments. An inhomogeneous distribution of strain and trion concentration was recently shown on \MoSe\ on Si/SiO$_2$ substrates\,\cite{JakubczykNL16}. We will report measurements on both samples in this paper.  

In order to measure the exciton density and dephasing dynamics, we perform three-pulse FWM spectroscopy. The excitation pulses are derived from a femtosecond Ti:Sapphire laser (Coherent Mira 900) with 76\,MHz repetition rate. The pulse spectra used are given in \Fig{fig:extinction}(b) and (c), and show a FWHM of 5.5\,meV and 11\,meV, corresponding to pulse durations of 240\,fs and 120\,fs, respectively. The pulses are focussed onto the sample to a diffraction limited spot of about 16\,$\mu$m intensity FWHM.  
The first pulse ($P_1$) induces a coherent polarization in the sample, which after a delay $\tau_{12}$ is converted into a density grating by the second pulse ($P_2$). The third pulse ($P_3$), delayed by $\tau_{23}$ from $P_2$, is diffracted by this density grating, yielding the FWM signal. In the employed heterodyne technique\,\cite{BorriBook09} the pulse trains are radio-frequency shifted resulting in a frequency-shifted FWM field which is detected by its interference with a reference pulse. In the investigated inhomogeneously broadened ensemble, the FWM signal is a photon echo emitted at a time $\tau_{12}$ after $P_3$, and the microscopic dephasing is inferred from the decay of the photon echo amplitude versus $\tau_{12}$. Conversely, the decay of the photon-echo amplitude versus $\tau_{23}$ probes the exciton density dynamics\,\cite{ShahBook96}. This setup was used in previous works\,\cite{MasiaPRL12, AccantoACSN12, NaeemPRB15}, and more details can be found in the supplement of \Onlinecite{NaeemPRB15}. For all the data shown in this work, each beam had equal power $P$. The polarization of the beams was controlled by waveplates. The polarization configuration of $P_1$, $P_2$, $P_3$ and detection will be abbreviated by four symbols in this order, e.g.  $(\rightarrow, \rightarrow,\uparrow,\uparrow)$ for cross-linear polarization of $P_{1,2}$ to $P_3$ and detection.
The repetition rate of the excitation corresponds to a period $\TREP$ of about 13\,ns, which limits the maximum relaxation time which can be extracted from the delay time dependence. Furthermore, the density modulation frequency of $(\Omega_1-\Omega_2)/(2\pi)=1\,$MHz provides a single pole high-pass filter for the response, with a cut-off at lifetimes of $1/(\Omega_1-\Omega_2)\sim 160$\,ns.

\section{\label{sec:dephasing} Exciton and Trion dephasing}

\begin{figure}[bt]
	\includegraphics[width=\columnwidth]{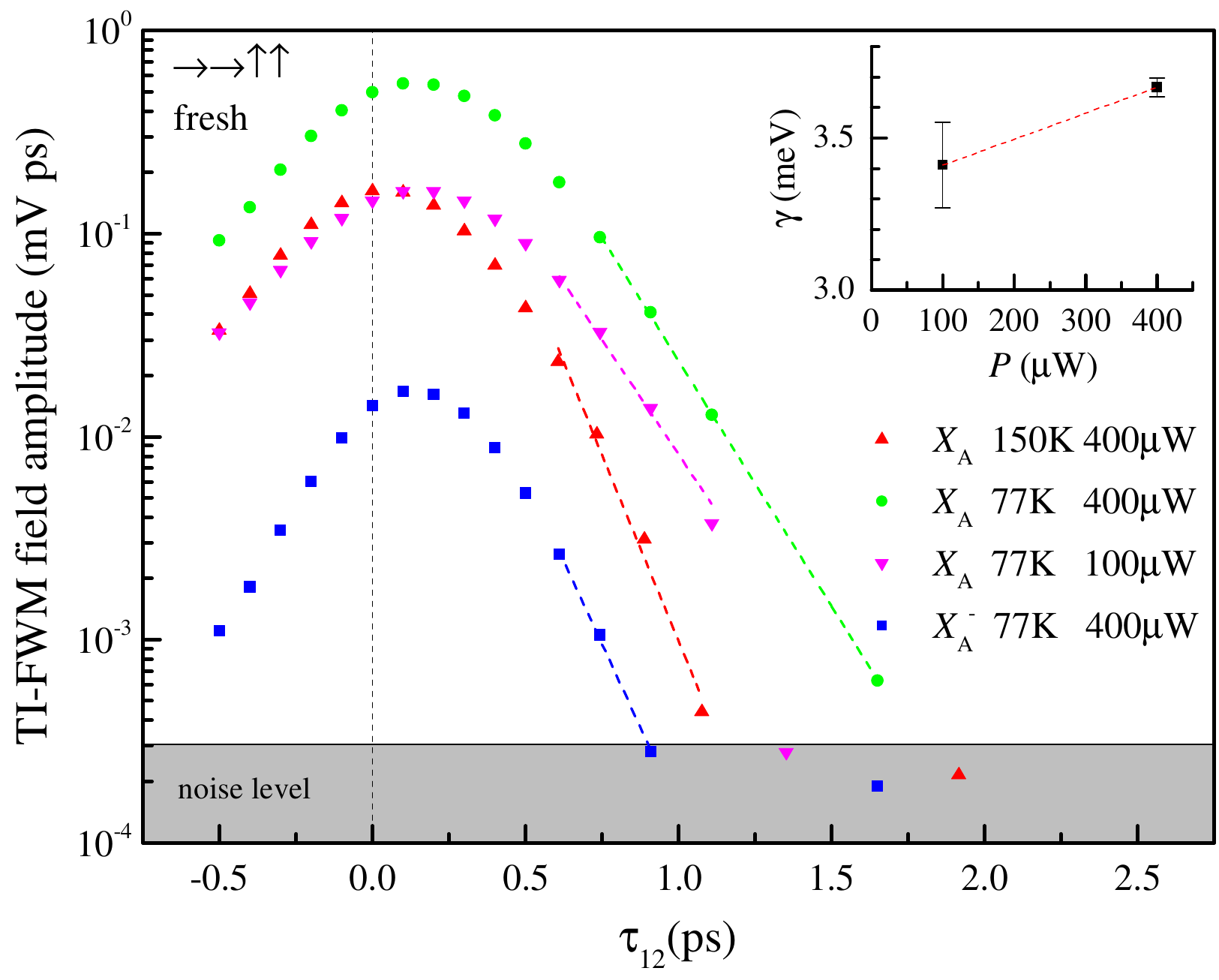}
	\caption{$\XA$ and $\XAT$ time-integrated four-wave mixing (TI-FWM) field amplitude as a function of $\tau_{12}$ for different temperatures and excitation powers and $\tau_{23}$ of 1\,ps and 1.2\,ps respectively, measured for the fresh sample. The dashed lines are exponential fits to the data. The inset shows the extracted power dependence of the $\XA$ homogeneous linewidth $\gamma$ at $T=77$\,K.}
	\label{fig:dephasing}
\end{figure}

The dephasing of $\XA$ and $\XAT$ was measured by FWM using $(\rightarrow, \rightarrow,\uparrow,\uparrow)$ polarization on the fresh sample. The measured photon echo amplitude versus  $\tau_{12}$ is shown in \Fig{fig:dephasing}. We can see a rapid decay of the echo amplitude with $\tau_{12}$. Since the laser pulse spectrum is narrower than the broadening of the transitions (see \Fig{fig:extinction}(b)), the photon echo duration is given by the laser pulse width. To extract the dephasing time $T_2$ we fit the data beyond pulse overlap $\tau_{12}>0.5$\,ps with an exponential decay $\propto\exp(-2\tau_{12}/T_2)$. From the fits, shown as dashed lines in \Fig{fig:dephasing}, we obtain for $\XA$ at T=77\,K a dephasing time $T_2$ of $(386\pm 16)$fs for P=100\,$\mu W$ and of $(359\pm 3)$fs for P=400\,$\mu W$. At 150\,K and for P=400\,$\mu W$ we obtain a $T_2$ of $(237 \pm 18)$fs. The data for 150\,K is close to the limit of the time-resolution of the experiment. The extracted exciton homogeneous linewidth $\gamma=2\hbar/T_2$ shows a weak power dependence (see inset of \Fig{fig:dephasing}), which we fitted using $\gamma=\gamma_0+\gamma_PP$. We find a zero-density homogeneous linewidth $\gamma_0$ of 3.3\,meV and $\gamma_P\approx 1\,$ meV/mW.

The exciton dephasing in \MoSe\ is affected by radiative decay, and scattering by phonons and charge carriers. The trion absorption found in \Sec{sec:sample} shows the presence of charge carriers, which create an additional dephasing and relaxation channel for excitons.
In \Onlinecite{AroraNS15}, the temperature dependent linewidth (FWHM) of the $\XA$ resonance in reflection was fitted with 
\be \gamma(T)=\gamma_0+\frac{A}{e^{\EA/\kB T}-1} \ee
using $A=72\,$meV and $\EA=30$\,meV. This width includes homogeneous and inhomogeneous broadening. The phonon activated term, representing a homogeneous broadening, amounts to $33$\,meV at $T=300\,$K, $7.8$\,meV at $T=150\,$K, and $0.79$\,meV at $T=77\,$K. In \Onlinecite{JakubczykNL16} an additional linear term of 0.03\,meV/K was found, and a homogeneous broadening of about 3.5\,meV at 77\,K, which is consistent with our finding.  

The trion $\XAT$ shows a faster dephasing than $\XA$, with $T_2=(266\pm13)$\,fs at $T=77$\,K, corresponding to a homogeneous broadening of $(5.0\pm0.3)$\,meV. This is attributed to the additional carrier, the charged character, and the lower binding energy, increasing the scattering rate.

\section{\label{sec:exciton} Exciton population dynamics}

In this section we discuss the measured FWM density dynamics as a function of $\tau_{23}$ for $\tau_{12}=0$\,ps after resonantly exciting A-excitons $\XA$. 

\subsection{\label{subsec:excitondensity} Density dependence}

\begin{figure}[bt]
	\includegraphics[width=0.62\columnwidth]{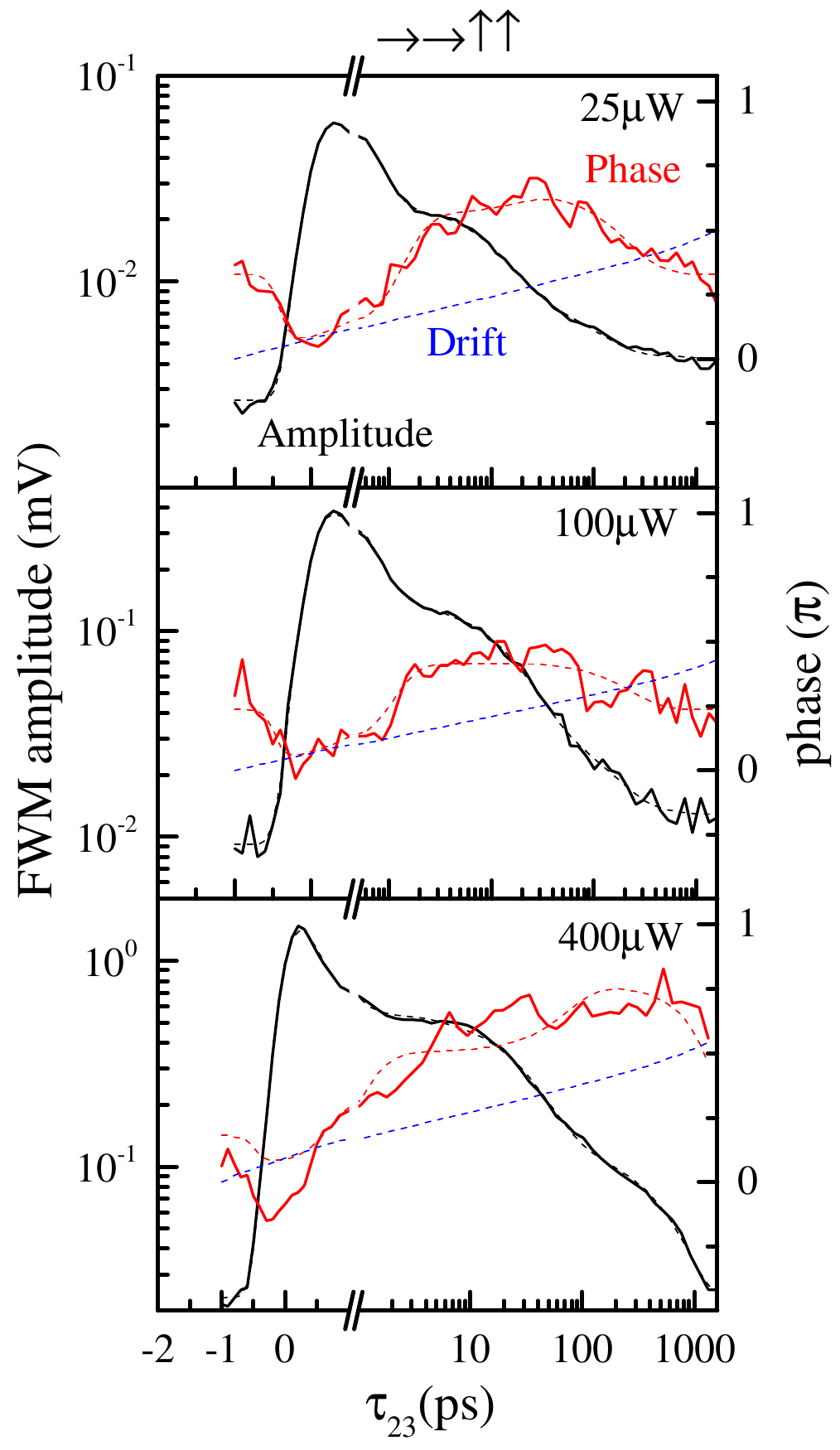}
	\caption{Density dynamics of the fresh sample. FWM field amplitude (black solid line) and phase (red solid line) as function of the delay $\tau_{23}$ for different excitation powers $P$ as given and cross-linear polarization configuration $(\rightarrow, \rightarrow,\uparrow,\uparrow)$. The dashed lines are fits to the data. The fitted phase drift (blue dashed line) has been subtracted from the data. Excitation resonant to $\XA$ at T=77\,K.}
	\label{fig:FWMDensityPowerFresh}
\end{figure}

\begin{figure}[bt]
\includegraphics[width=\columnwidth]{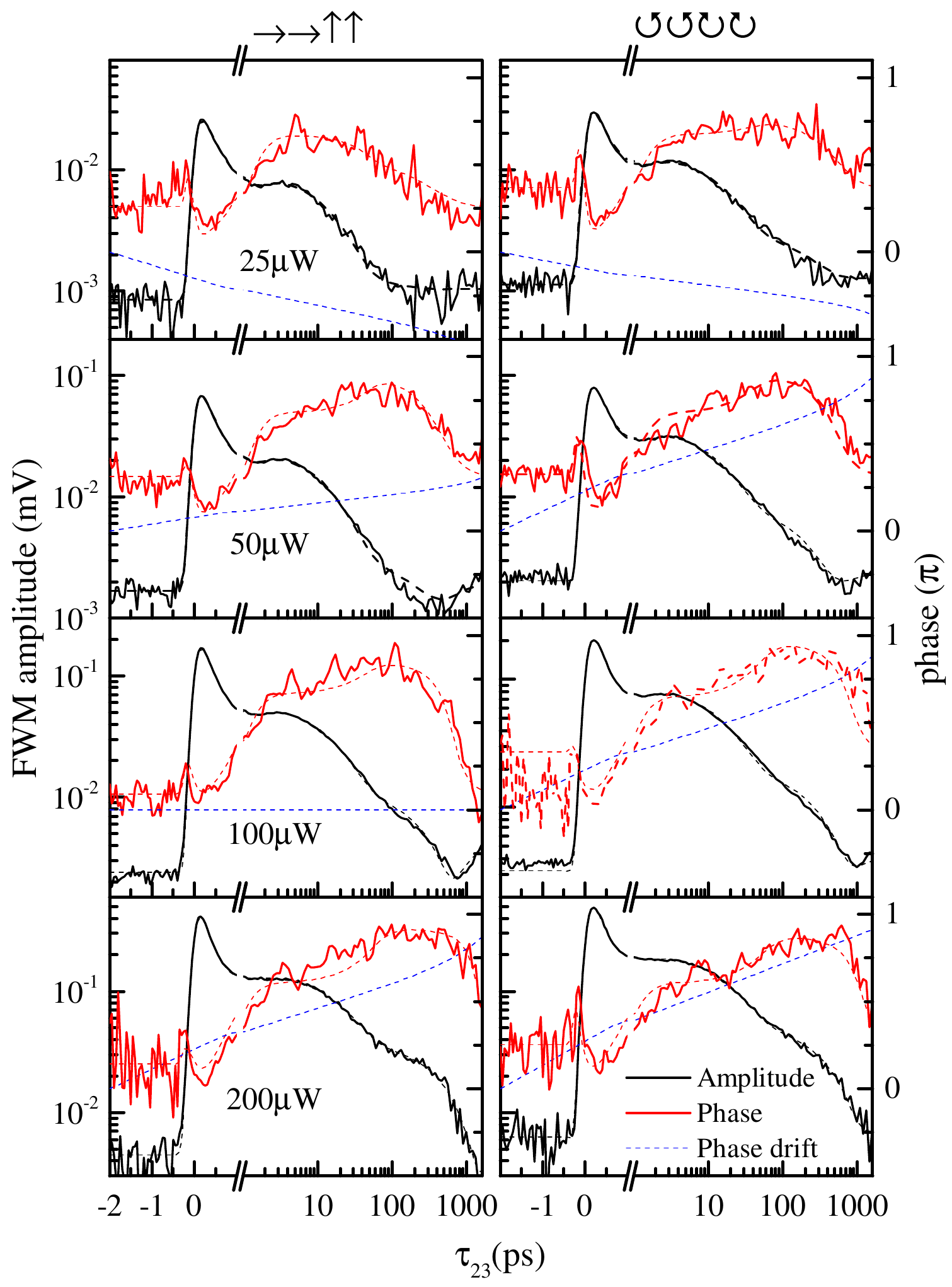}
	\caption{As \Fig{fig:FWMDensityPowerFresh}, but for the aged sample and two polarization cofigurations, cross-linear $(\rightarrow, \rightarrow,\uparrow,\uparrow)$ and cross-circular $(\circlearrowleft\circlearrowleft\circlearrowright\circlearrowright)$.}
	\label{fig:FWMDensityPower}
\end{figure}

The measured FWM field amplitude and phase for different excitation power $P$ are shown in \Fig{fig:FWMDensityPowerFresh} for the fresh sample and in \Fig{fig:FWMDensityPower} for the aged sample for linear and circular polarization configurations over 4 orders of magnitude in delay $\tau_{23}$. We find processes with time scales from the picosecond to the nanosecond range. To quantitatively analyze the data, we use a fit with the complex multi-exponential response function
\be
R(\tau)\propto \ATPA e^{i\varphiTPA}\delta(\tau) + \sum_{n} A_n\theta(\tau)\exp\left(i\varphi_n-\frac{\tau}{\tau_n}\right)
\label{ComplexResponse}
\ee
where $A_n$, $\varphi_n$, and $\tau_n$ are amplitude, phase, and decay time of the $n$-th decay process and $\ATPA$ is a non-resonant instantaneous component such as Kerr effect or two-photon absorption. \Eq{ComplexResponse} describes the FWM complex field as a coherent superposition of exponential decay components created by the pump. Using a complex fit takes into account the  different phases of the components, given by the relative effect on absorption and refractive index. The measurements, see for example $P$=100\,$\mu$W in \Fig{fig:FWMDensityPower}, show a characteristic change of the phase versus time -- the phase at negative delays is similar to the phase of the resonantly generated signal after pulse overlap, at $\tau_{23}>100\,$fs. This component decays in the first picoseconds after which a phase change of about $\pi/2$ is observed together with a slower dynamics. After around 20\,ps a second a bit smaller phase shift is seen, and later, around 1\,ns, the phase starts to shift back to its initial value. The measured phase contains a slow drift, which is caused by temperature drift of the setup, changing the relative phase of reference and probe pulses over the time $t$ of the measurement of a few minutes. This phase drift has been included in the fitting procedure in a prefactor $\exp(i(\varphi_0+\varphi_0't))$ using a linear time dependence. In the fit, four components $n=1..4$ are used. We fix the phase of the fastest component to $\varphi_1=0$ so that the phases of the other components represent the relative phases to the first component. The response is convoluted with a periodic Gaussian to reflect the excitation pulses of full-width at half-maximum (FWHM) $2\sqrt{\rm ln2}\tau_0$ in amplitude, and repetition period $\TREP$, yielding the fit function
\begin{align}
F(t,\tau) = \exp(i(\varphi_0+\varphi_0't))\left\{ \ATPA\exp\left(i\varphiTPA-\frac{\tau^2}{\tau_0^2} \right) \right.\nonumber \\+\sum_{n} A_n \left[\frac{1}{e^{\frac{\TREP}{\tau_n}}-1}+ \frac{1}{2}\left(1+{\rm erf}\left(\frac{\tau}{\tau_0}-\frac{\tau_0}{2\tau_n}\right)\right)\right] \nonumber\\
\left.\times\exp\left(i\varphi_n+\frac{\tau_0^2}{4\tau_n^2}-\frac{\tau}{\tau_n}\right)\right\}
\label{eqn:ComplexFit}
\end{align}
which includes the pile-up of signal due to the finite $\TREP$. The resulting fits are given in \Fig{fig:FWMDensityPowerFresh} for the fresh and \Fig{fig:FWMDensityPower} for the aged sample. In the figure, we show for clarity the phase corrected for the phase drift, and the phase drift separately.

\begin{figure}[h!]
	\includegraphics[width=\columnwidth]{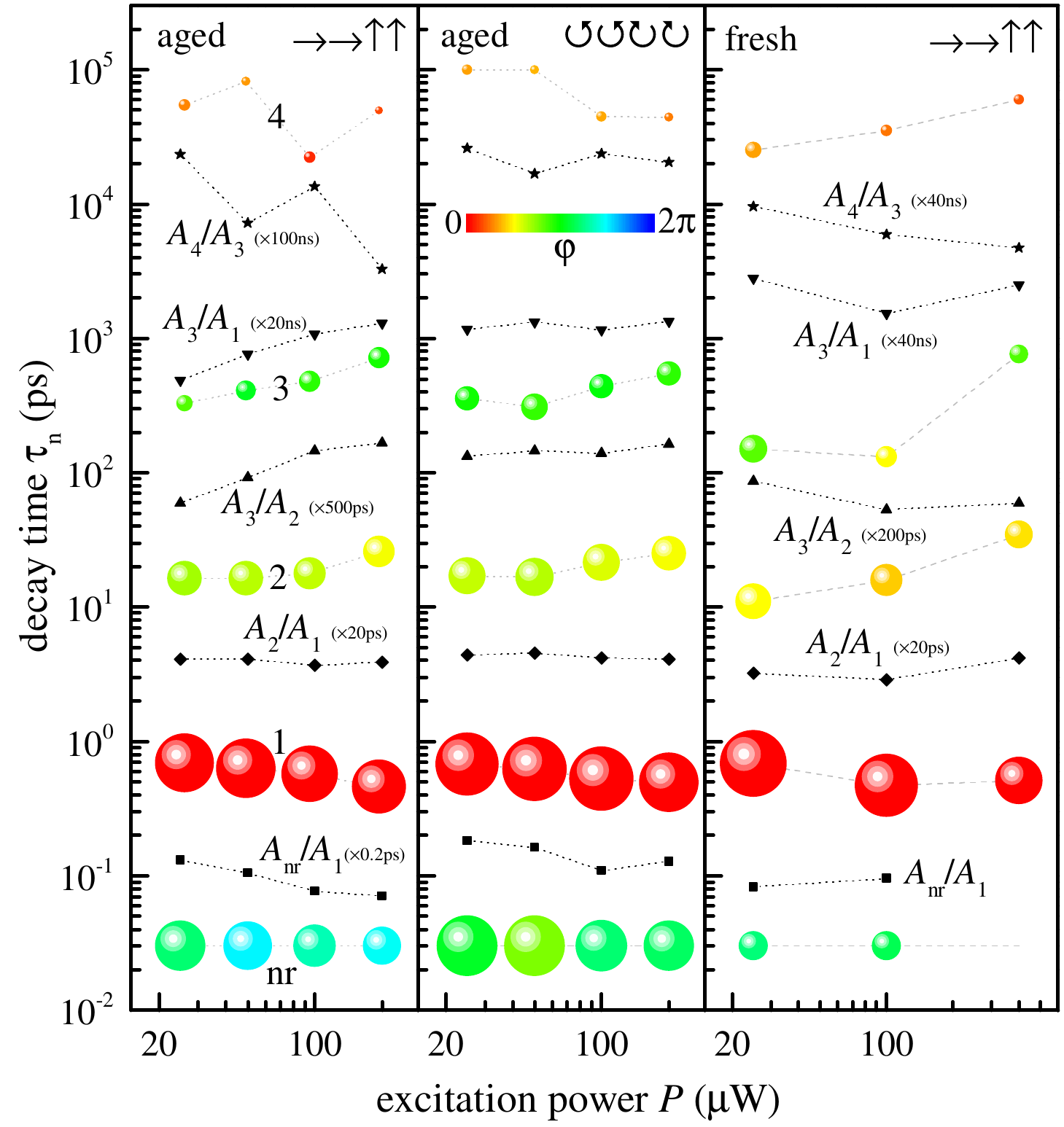}
	\caption{Results of the fit of \Eq{eqn:ComplexFit} to the data shown in \Fig{fig:FWMDensityPowerFresh} and \Fig{fig:FWMDensityPower}. The amplitudes $A_n$ normalized to the third-order scaling $P^{3/2}$ with the excitation power are given by the volume of the symbol (diameter proportional to $A_n^{1/3}P^{-1/2}$, and for absolute scaling $A_1=40\,\mu$V at $P=25\,\mu$W for the aged  $(\rightarrow, \rightarrow,\uparrow,\uparrow)$ sample, and $80\,\mu$V for the fresh sample). The phases $\varphi_n$ are given by the symbol color according to the colormap. $\ATPA$ is shown at a decay time of 30\,fs. Selected amplitude ratios, scaled as labelled, are given as symbols. Dotted lines are guides for the eye.}
	\label{fig:FitParaPower}
\end{figure}

The decay times $\tau_n$, amplitudes $A_n$ and phases $\varphi_n$ resulting from the fit to the complex data are shown in \Fig{fig:FitParaPower}. The amplitudes are shown normalized to the scaling $P^{3/2}$ expected for a $\chi^{(3)}$ process. The fastest component with $\tau_1\sim 0.6$\,ps represents the dominating amplitude $A_1$, about 5 times larger ($A_1/A_2 \sim 5$) than the second component with $\tau_2\sim 20$\,ps and a phase of $\varphi_2\sim \pi/2$. The next component with $\tau_3\sim 400$\,ps has an amplitude $A_3$ which is again about 4 times lower than $A_2$, ($A_2/A_3 \sim 4$), and a phase of $\varphi_3 \sim \pi$, i.e. it is out of phase to the first component. The fourth component with $\tau_4\sim 20..100$\,ns has an amplitude which is about 10 times lower than $A_3$, ($A_3/A_4 \sim 10$). It is longer than $\TREP$, and is responsible for the signal at negative delay due to a pile-up of previous repetitions. It has a phase $\varphi_4\sim 0..\pi/4$, i.e. similar to the initial component.

We observe a weak dependence of the dynamics on the excitation power. With $P$ increasing over an order of magnitude, $\tau_1$ is decreasing by a few 10\%, and $\tau_2$ and $\tau_3$ are increasing by about a factor of 2. The dominant amplitude $A_1$ is scaling as expected for a third-order process, as seen by the near constant symbol size versus $P$, apart from the highest powers used. The ratio $A_2/A_1$ does not change with $P$, but the ratio $A_3/A_2$ is increasing by a factor of 2 for the cross-linear aged case, near constant for the cross-circular aged case and slightly decreasing for the cross-linear fresh case. For the longest component the values of $\tau_4$ and $A_4$ are partially correlated since the signal measured at negative delay is given by a combination of both, and at the maximum positive delay of 1.6\,ns a part of component 3 still is remaining.    
\begin{figure}[b]
	\includegraphics[width=\columnwidth]{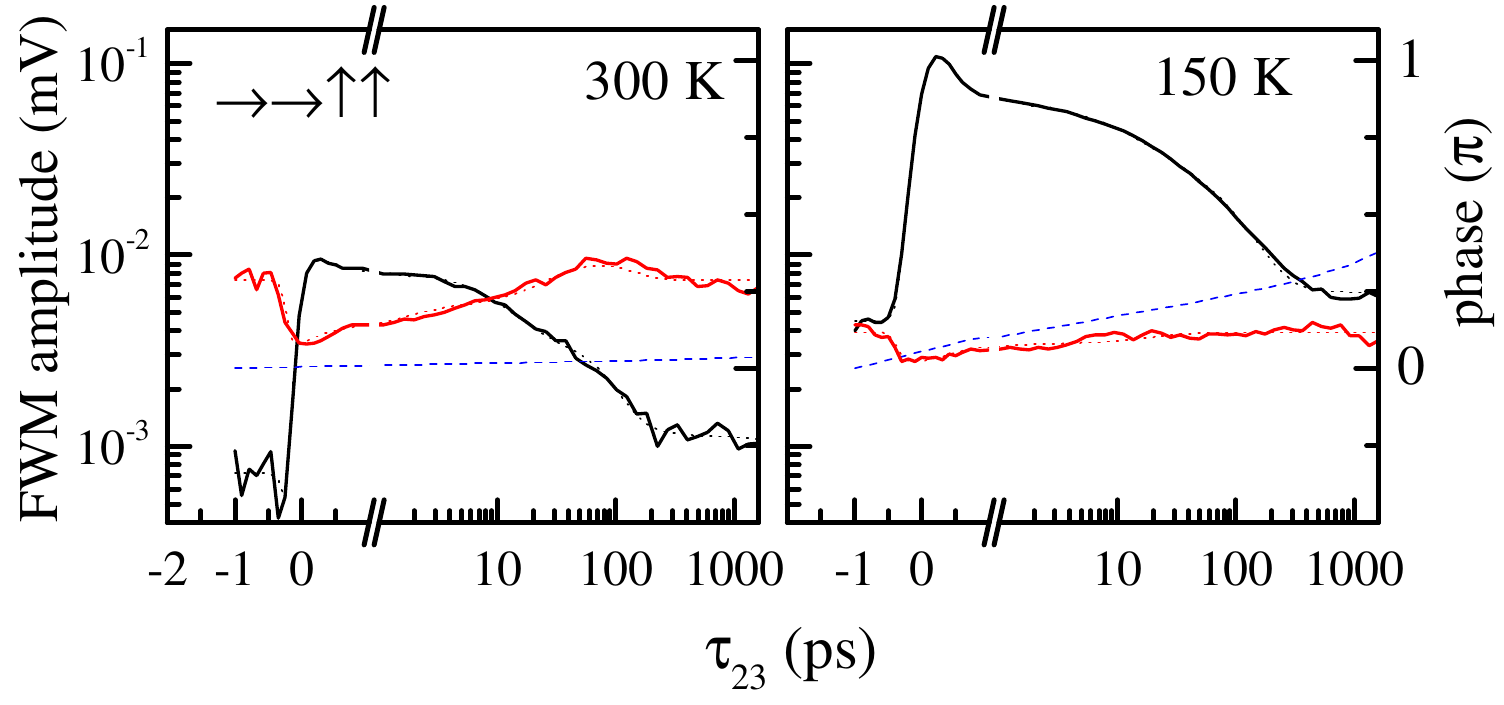}
	\caption{Density dynamics of the fresh sample for different temperatures as indicated. FWM field amplitude and phase as function of the delay $\tau_{23}$ for $P=100\,\mu$W, cross-linear $(\rightarrow \rightarrow\uparrow\uparrow)$ polarization configuration, and excitation resonant to $\XA$.}
	\label{fig:FWMDensityTemperature}
\end{figure} 

\subsection{\label{subsec:excitontemp} Temperature dependence}   

To investigate the influence of phonon-scattering and thermal distribution across states, we have measured the density dynamics for different temperatures. Additionally to the $T=77$\,K data shown in \Fig{fig:FWMDensityPowerFresh}, for which the thermal energy $\kB T$ is about 6.6\,meV, we have taken data for $150\,$K and $300\,$K, corresponding to $\kB T$ of 13\,meV and 26\,meV, respectively. A power $P=100\,\mu$W was used, allowing for sufficient dynamic range while limiting the density induced effects. The resulting dynamics is given in \Fig{fig:FWMDensityTemperature}. The excitation pulse center wavelength was shifted to match the $\XA$ for each temperature, compensating the temperature dependence of the band-gap.  

\begin{figure}
	\includegraphics[width=0.8\columnwidth]{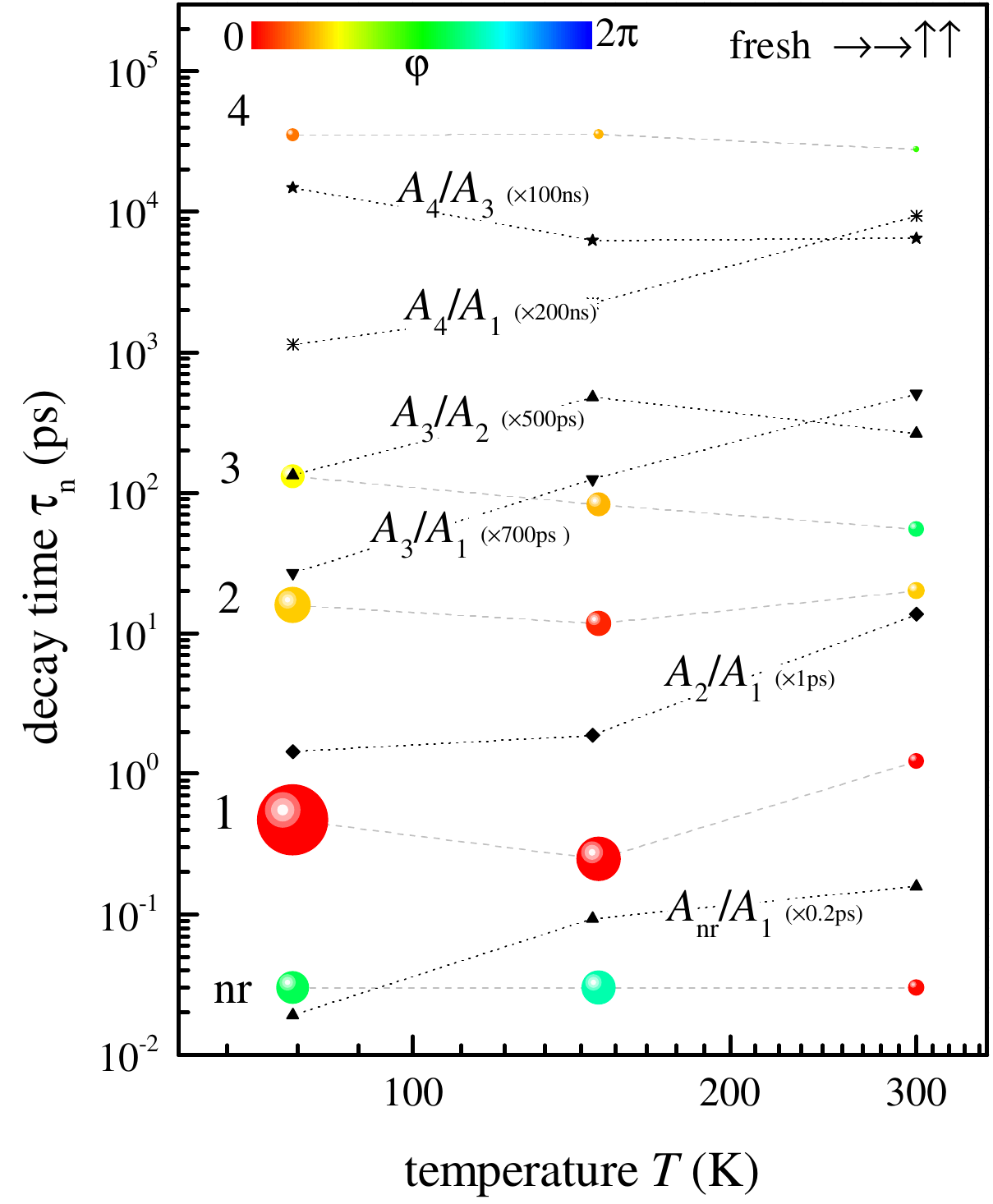}
	\caption{As \Fig{fig:FitParaPower}, but for the fresh sample only, as function of temperature for $P=100\,\mu$W and cross-linear $(\rightarrow \rightarrow\uparrow\uparrow)$ polarization configuration.}
	\label{fig:FitParaTemperature}
\end{figure}

We find that with increasing temperature, the main effect is a reduction of the decay in the first picosecond, while the subsequent dynamics is changing in a more subtle way. The results of a fit of the data with the multi-exponential decay model are given in \Fig{fig:FitParaTemperature}.   
We see the reduction of $A_1$ with increasing temperature as expected, and also an acceleration of the third component from $\tau_3\sim 130\,$ps at 77\,K to 55\,ps at 300\,K.
 
\subsection{\label{subsec:excdiscussion} Discussion}

To interpret the results, we remind ourselves that the strong exciton binding energy of a few hundred meV, combined with resonant excitation of the lowest exciton $\XA$, leads to a dominating role of excitons in the carrier dynamics. Free electron-hole pairs are expected to play a minor role, even at room temperature. Furthermore, for the same reason, the lower spin-split valence band, with a splitting energy of about 200\,meV, involved in $\XB$, is also expected to be not relevant. In the presence of extrinsic carriers, due to defects or doping, the resulting electron or hole density leads to the formation of trion states. These will be considered later.  
\begin{figure}
	\centering
	\includegraphics[width=8cm]{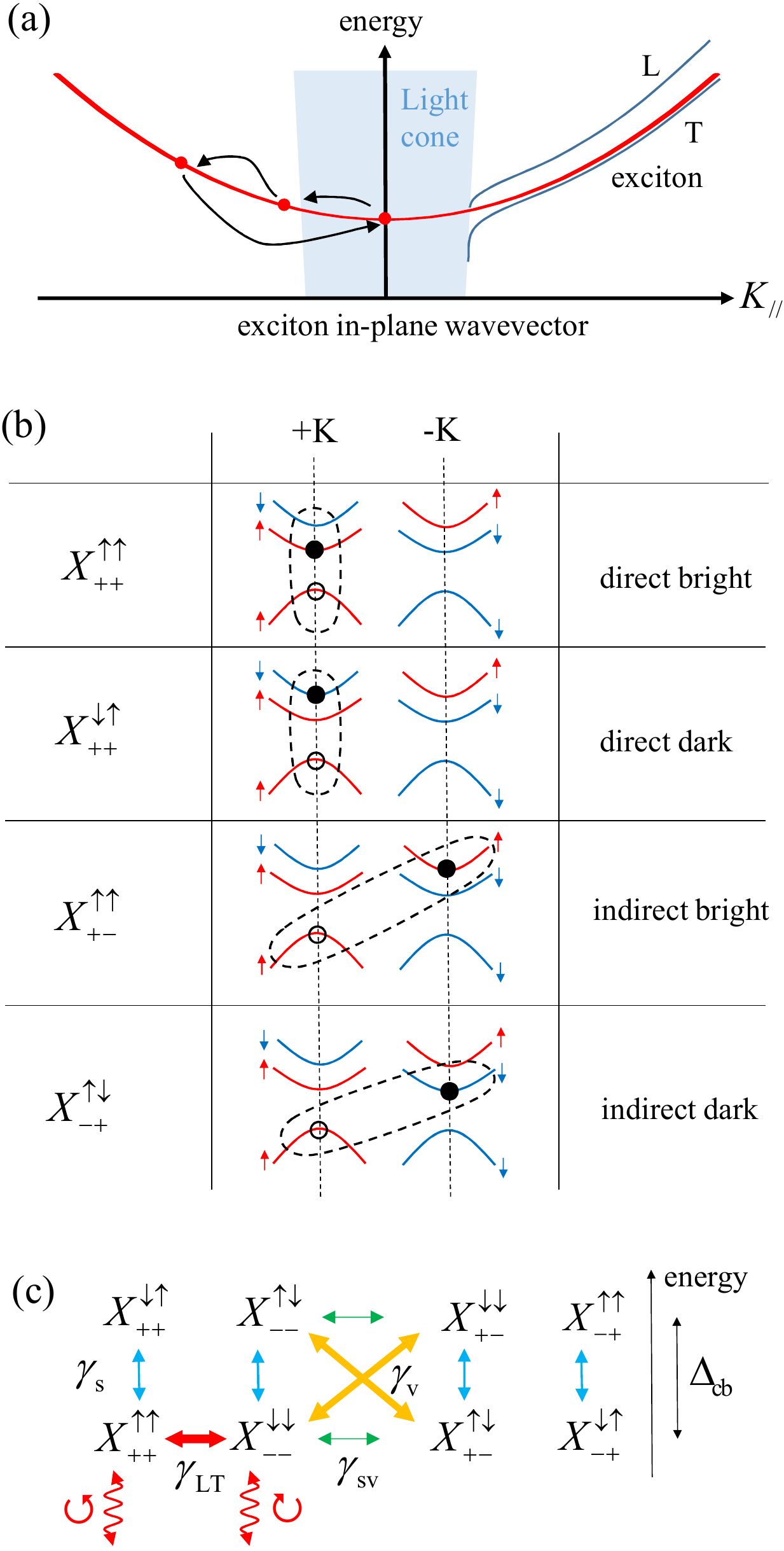}
	\caption{(a)\,Sketch of the exciton center of mass dispersion, and phonon-assisted scattering between inside and outside the light cone; (b)\,Possible electron-hole combinations in the single-particle electronic band structure forming the exciton. An occupied (unoccupied) electron state is represented by full (empty) black circle, with a red/blue vertical arrow indicating the corresponding electron spin. The nomenclature and properties of each exciton state is given. The lower energy valence band with opposite spin is not shown due to its large energy splitting of about 200\,meV. Each state shown has degenerate partner with inverted indices, e.g. $X^{\uparrow\uparrow}_{++}$ and $X^{\downarrow\downarrow}_{--}$.}
	\label{fig:StateSketch}
\end{figure}
The remaining intrinsic states expected to be relevant to the exciton dynamics are sketched in \Fig{fig:StateSketch}. Firstly, the excitons have an in-plane dispersion (see \Fig{fig:StateSketch}a) given by their center-of-mass (COM) mass $M^{\ast}$, the sum of electron and hole effective masses, and $M^{\ast}\sim\,m_0$\,\cite{HorzumPRB13}. We note here that the prediction\,\cite{YuNC14} of a Dirac-cone type exciton dispersion in the light cone is disputed\,\cite{GartsteinPRB15,WangPRB16}. Neglecting in-plane disorder, the exciton states are eigenstates of the in-plane wavevector $\bK$. They couple to external propagating light only if they are ``slow'', having a $\bK$ within the light cone, i.e. $|\bK|<K_0=n\omega/c$, where $n$ is the refractive index of the surrounding medium, as is well established for quantum well excitons\,\cite{AndreaniPRB90, AndreaniBook94}. We will use here $n=1.5$ from the quartz substrate and superstrate. A similar value is obtained considering instead \MoSe\ sandwiched between hBN with $n \sim 2$ and vacuum. Excitons having a $\bK$ outside the light cone, which we call ``fast'' here, couple to evanescent fields which results in a polariton energy shift\,\cite{GartsteinPRB15} of similar magnitude as the radiative linewidth inside the light cone, as shown in the right side of the sketch. The polariton interaction creates exciton eigenstates which are linearly polarized along or orthogonal to $\bK$, called longitudinal (L) and transversal (T) exciton-polaritons. The exciton radiative lifetime within the radiative cone $\taur$ scales with the square of the exciton oscillator strength, and is about 10\,ps in the classical 2D system - GaAs quantum wells\,\cite{AndreaniBook94}. The large exciton binding energy in \MoSe\ leads to a faster lifetime, in the sub-picosecond regime\,\cite{DufferwielNC15,WangPRB16, JakubczykNL16, RobertARX16}.
This situation is similar to the fast radiative lifetimes in the 1\,ps range observed in two-dimensional CdSe platelets\,\cite{NaeemPRB15}, which have an exciton binding energy of 100-300\,meV. In presence of in-plane disorder, the eigenstates have a finite width in $\bK$, yielding longer minimum radiative lifetimes, as only part of the localized exciton wavefunction is within the radiative cone\,\cite{SavonaPRB06}.
Scattering between slow and fast excitons can occur by interaction with phonons, excitons, and charge carriers. Assuming thermal equilibrium within the exciton dispersion, one can define\,\cite{AndreaniBook94, WangPRB16} an effective radiative decay time of 
\be 
\tau(T)=\frac{3\kB T}{4E_0}\taur\,, \; \mbox{with}\; E_0=\frac{\hbar^2K_0^2}{2M^*}
\ee
with the exciton kinetic energy at the edge of the light cone $E_0=4.5\,\mu$eV. Note that in this expression the polariton shift and local disorder effects are neglected, which should be a reasonable approximation for $\kB T$ much larger than the relevant energy scales of a few meV. This expression yields the values $\tau(77\,K)\approx 1100\,\taur\approx 440\,$ps, $\tau(150\,K)\approx 2150\,\taur\approx 860\,$ps, and $\tau(300\,K)\approx 4310\,\taur\approx 1.7\,$ns, using\,\cite{DufferwielNC15} $\taur=0.4$\,ps. The assumed thermalization in this model requires that the scattering is much faster than $\taur$, which is not the case at $T=77$\,K for which we measure a dephasing time of about 0.36\,ps (see \Sec{sec:dephasing}). We therefore expect at this temperature a two component decay, first the fast radiative decay of the initially excited slow excitons, and later a decay of the fast excitons, via scattering to the slow excitons and subsequent radiative emission. The resulting decay rate is lower than predicted in the thermalized case, due to the sub-thermal population of states inside the radiative cone. At room temperature instead, the phonon scattering is likely to be sufficiently fast to allow thermalization. Accounting for disorder, leading to in-plane localization, the maximum radiative rate of the exciton states decreases\,\cite{ZimmermannBook03, SavonaPRB06}, allowing for better thermalization. The effect of inhomogeneous broadening on the dephasing has recently been shown\,\cite{JakubczykNL16}. This broadening will depend on the specific sample investigated, as it is related to the influence of the embedding layers. 

Further to the in-plane dispersion, the exciton has different spin and valley states\,\cite{XiaoPRL12,DeryPRB15}, as sketched in \Fig{fig:StateSketch}b. We denote these states as $X_{\ie \ih}^{\se \sh}$ where $\ie$ and $\ih$ are the valley index $+$ or $-$, and $\se$ and $\sh$ are the spin $\uparrow$ or $\downarrow$, of the electronic state, of electron and hole, respectively. The optical transitions are circularly polarized according to the valley index, and conserve spin and valley index. We call an exciton state bright for $\se=\sh$, and dark otherwise. Furthermore, an exciton which has equal valley index $\ie=\ih$ is called direct, and indirect otherwise. The direct bright states $X_{++}^{\uparrow \uparrow}$ and $X_{--}^{\downarrow \downarrow}$ couple to external light and thus have a radiative decay, and the polariton features discussed above. Notably, the polariton self-energy is coupling these two states to form linearly polarized superpositions L and T, with a splitting of a few meV for the fast excitons\,\cite{GartsteinPRB15}. This splitting corresponds to an oscillation between $X_{++}^{\uparrow \uparrow}$ and $X_{--}^{\downarrow \downarrow}$ on the picosecond timescale, which can explain the observed fast spin relaxation\,\cite{WangAPL15}, and is indicated by $\gLT$. Note that the eigenstates and splitting depend on $\bK$, so that instead of a beating a fast decay can be expected.  Including in-plane disorder, the LT splitting is transformed into a fine-structure splitting of the disorder-localized exciton states, providing a fast spin relaxation without dephasing\,\cite{ZimmermannPE01}. 

The direct dark states $X_{++}^{\downarrow \uparrow}$ and $X_{--}^{\uparrow \downarrow}$ are assumed to have higher energy  due to the conduction band splitting of about\,\cite{DeryPRB15} $\Dc=20\,$meV. The direct dark states $X_{++}^{\uparrow \downarrow}$ and $X_{--}^{\downarrow \uparrow}$ are not considered in this discussion due to the large valence band splitting. The electron-hole exchange interaction is expected to lead to an energy splitting of the degenerate direct bright and indirect dark excitons, but reliable measurements or calculations of this effect are not available to our knowledge.  

Scattering between the different states can occur via phonon absorption or emission. The phonon band structure is given in \Onlinecite{HorzumPRB13}. A sketch of the different scattering rates considered is given in \Fig{fig:StateSketch}c. One can distinguish scattering changing the valley only, denoted by $\gv$, scattering changing the spin only, denoted by $\gs$, and scattering changing both valley and spin, $\gvs$. 

$\gvs$ connects the direct bright and indirect dark state, which have, apart from possible exchange energies, the same energy. Since the phonon has to provide the wavevector difference between the valleys, which is given by the $K$ point of the lattice, there are no suited energy and momentum conserving phonon modes available and we would thus expect that $\gvs$ is small. $\gs$ connects direct bright and dark states, which are split by $\Dc$, but have equal wavevector. This scattering process can occur via $\Gamma$ point phonons of similar energy, such as the $E''$ or $A'$ modes. However, phonon interactions couple to the orbital part of the wavefunctions, and spin can only be flipped via spin-orbit interaction. We are not aware of explicit calculations of the related scattering rates. $\gv$ connects the direct bright with the indirect bright state, and the direct dark with the indirect dark state. Both have the energy difference $\Dc$, and a momentum difference between the -K and +K point. Phonon modes matching the energy and momentum required for these transitions are present, such that we expect $\gv$ to be the largest of all three rates. 

In view of these properties of the excitonic states in \MoSe\, we now discuss the observed FWM density dynamics, starting with $T=77$\,K. The optical excitation populates the direct bright exciton close to the center of the radiative cone. For linear polarization, a superposition of $X_{++}^{\uparrow \uparrow}$ and $X_{--}^{\downarrow \downarrow}$ is populated, while for circular polarization, only $X_{++}^{\uparrow \uparrow}$ is populated.
In both cases, the states are expected to show a fast radiative recombination, consistent with the first component in \Fig{fig:FitParaPower}. Its decay time of about 0.6\,ps, observed for both polarizations, is consistent with the expected radiative lifetime. Importantly, a scattering towards other states would not be expected to change the signal so strongly, as the exciton-exciton interaction is not expected to be strongly affected,  considering that the involved exciton kinetic energies are much smaller than the exciton binding energy. This scattering is therefore not expected to be prominent in the FWM signal. A competing process to radiative decay could be the phonon-assisted formation of a trion. The slightly decreasing $\tau_1$ with power could be due to exciton-exciton scattering into the fast states or other spin-valley states. 

Once the initially excited population of slow bright excitons is gone, the remaining fast excitons would be expected to have an effective radiative decay at least thousand times slower, as calculated earlier, in the order of 1\,ns. The fast excitons show, as discussed before, a large LT splitting, which is mixing the $X_{++}^{\uparrow \uparrow}$ and $X_{--}^{\downarrow \downarrow}$ states. The memory of the initially excited state, which is different for circularly and linearly polarized excitation, is therefore lost in the fast direct states within the first picoseconds, and the subsequent dynamics is essentially independent of the excitation polarization configuration, consistent with our observations, see e.g. \Fig{fig:FWMDensityPower}. The dynamics given by the second component with times $\tau_2$ of about 20\,ps is therefore attributed to scattering of the fast bright states into the indirect dark states, possibly via the indirect bright states involving the rates $\gv$ and $\gs$, the latter being the smaller rate is expected to dominate the observed timescale. The resulting change of their interaction with the probed slow direct bright states changes the phase of the signal. Notably, the indirect bright and direct dark states at about 20\,meV higher energy will not carry a significant population at T=77\,K, considering the Boltzmann factor $\exp(-20/6.6)\approx 0.05$ - they are thus only intermediate states facilitating the transfer between direct bright and indirect dark state.  

Once the population of the fast direct bright exciton states and indirect dark excitons have thermalized by this process, the exciton density is decaying by scattering back to the direct bright state, or non-radiative processes. This process is attributed to the third component with $\tau_3\approx 300$\,ps.
The different FWM signal phases of the components can accordingly be attributed to the somewhat different exciton-exciton interactions with the slow direct bright states, which are carrying the FWM polarization. Component 1 is due to the interaction within these states, component 2 due to the interaction with fast direct bright states, and component 3 due to the interaction with indirect dark excitons.
Notably, the amplitudes of the components are scaling approximately with their decay rate, as would be expected for this picture of three exciton reservoirs with bidirectional scattering.

The increasing amplitude of the third process with excitation power could accordingly be related to exciton-exciton scattering from the direct bright to the indirect dark state, which requires exchange of electron or hole in the process $X_{++}^{\uparrow \uparrow}+X_{--}^{\downarrow \downarrow} \longrightarrow X_{+-}^{\uparrow \downarrow}+X_{-+}^{\downarrow \uparrow }$. This could increase also the radiative decay by a faster scattering into the slow bright excitons. 

The weak fourth component with an amplitude $A_4$ around 1\% of $A_1$ has a time-constant in the tens of nanoseconds and decreases its relative amplitude with increasing fluence. It could be related to thermalization of existing unpaired charge carriers after the optical excitation.      
   
Lets us now interpret the temperature dependence. With increasing temperature, the occupation of the phonons creating the scattering $\gs$ and $\gv$ is increasing according to their Bose - distribution. We can therefore expect that at 150\,K and 300,K, these two rates are dominating the dynamics, providing a faster thermalization between the various exciton states.

At 150\,K, this leads to an increased amplitude and reduced decay time of the third component. At 300\,K, the overall signal is reduced, due to the homogeneous broadening of the exciton of about 40\,meV, superseeding the inhomogeneous broadening (see also \Fig{fig:extinction}). This also means that the slow direct bright excitons are scattered within the pulse duration towards the other exciton states, and the dynamics is dominated by a single decay time around 30\,ps. At this temperature, the mobility of excitons in the sample could also allow the excitons to recombine non-radiatively at defects. However, in \Onlinecite{GoddePRB16}, a similar sample did not show a strong variation of the non-resonantly excited PL intensity $T>100$\,K, indicating that diffusion to defects and subsequent non-radiative decay is not significant.      
 
\subsection{\label{EEA}Exciton-exciton annihilation}

\begin{figure}
	\includegraphics[width=\columnwidth]{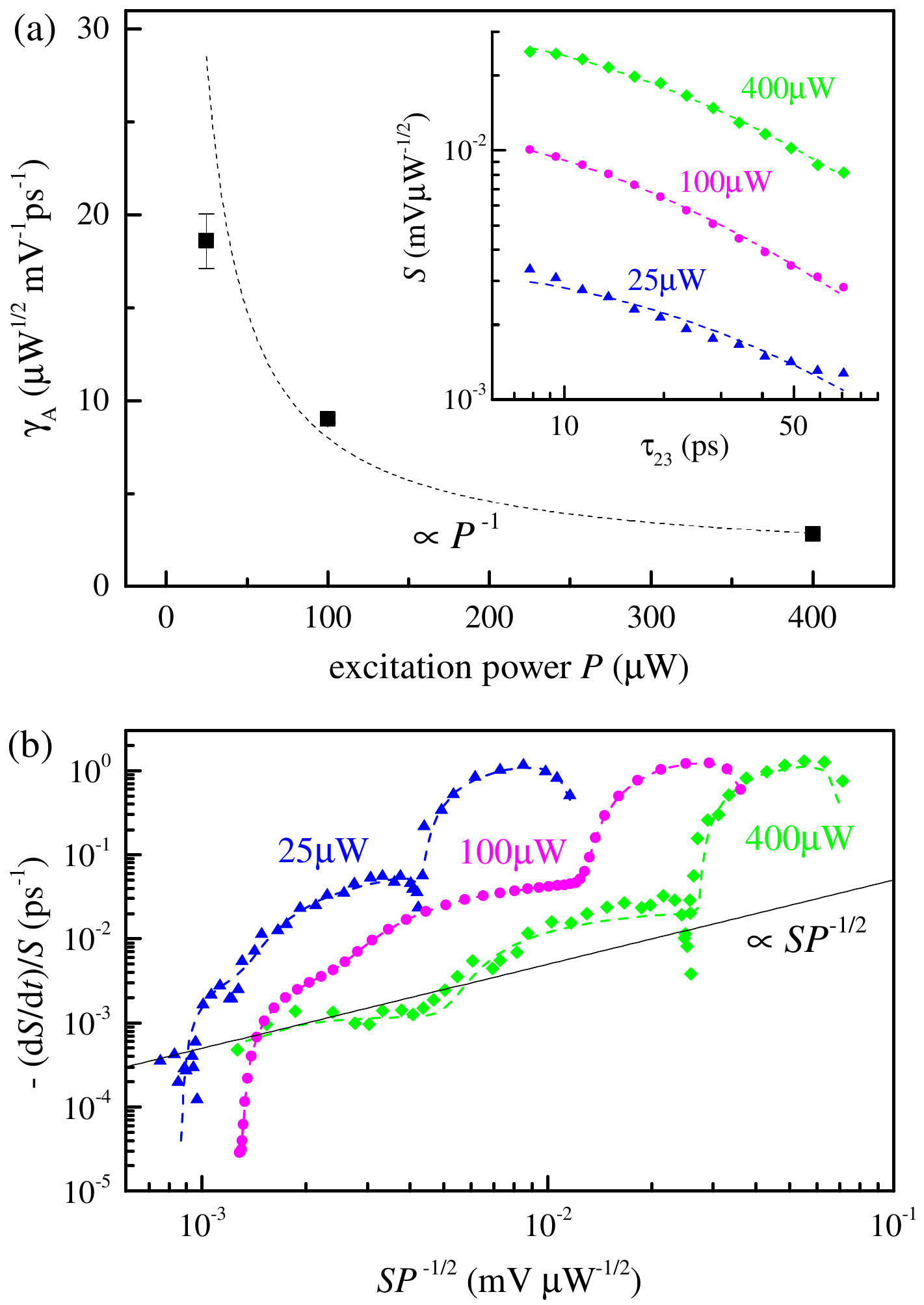}
	\caption{Exciton-exciton annihilation analysis. (a) Inset: Normalized FWM field amplitude $S/\sqrt{P}$ (symbols) as function of $\tau_{23}$ for different powers $P$ as indicated, and fits (dashed lines) according to \Eq{eqn:EEASolution}. Main: $\gEEA$ values (squares) determined by the fit; dashed line: $P^{-1}$ dependence. (b)\, Logarithmic decay rate of the FWM field amplitude $S$ versus $S/\sqrt{P}$ for different $P$ as indicated. Symbols: Measurements for $\tau_{23}\geq 0.3$\,ps; dashed lines: Fits shown in \Fig{fig:FWMDensityPowerFresh}; The solid line shows a dependence proportional to $S/\sqrt{P}$.}
	\label{fig:EEA_Combined_Fresh}
\end{figure}

Observation of exciton-exciton annihilation (EEA) in SL-TMD has been proposed to interpret carrier dynamics in $\text{WSe}_2$\,\cite{MouriPRB14}, $\text{MoS}_2$\,\cite{SunNL14} and $\text{MoSe}_2$\,\cite{KumarPRB14}, using largely varying EEA rates, and rates affected by the substrate\,\cite{YuPRB16}, while there are also reports where EEA was not invoked\,\cite{ShiACSN13,WangNL15}. In all these works, non-resonant optical excitation leading to free electron-hole pairs has been used. To evaluate whether EEA has a role in the dynamics we measure, we follow the analysis in \Onlinecite{KumarPRB14} using the same delay range from 5\,ps up to 60\,ps. The EEA rate equation describing the density decay is given by
\be
\frac{dN}{dt}=-\frac{N}{\tau}-\frac{\gEEA}{2} N^2
\label{eqn:EEAEquation}
\ee
with the density of excitons $N$, the low-density decay rate $\tau$, and the EEA rate $\gEEA$. In our experiments the exciton density is expected to be proportional to $S/\sqrt{P}$ where $S$ is the FWM field amplitude. We therefore fit this quantity as shown in the inset of \Fig{fig:EEA_Combined_Fresh}\,(a) for delays $\tau_{23}$ from 5\,ps to 55\,ps with the solution of \Eq{eqn:EEAEquation} neglecting the low density decay rate, given by\,\cite{KumarPRB14Note}
\be
N(t)=\left(N_0^{-1}+\gEEA t/2\right)^{-1}
\label{eqn:EEASolution}
\ee 
where $N_0$ is the initial exciton density. The resulting rates $\gEEA$ are given in \Fig{fig:EEA_Combined_Fresh}\,(a). We find that $\gEEA$ is strongly varying, being approximately inversely proportional to the power $P$. To avoid the approximation of negligible low density rate used in \Onlinecite{KumarPRB14}, we divide \Eq{eqn:EEAEquation} by $-N$, resulting in a linear dependence on $N$. We show the measured $-(dS/dt)/S$ as function of $S/\sqrt{P}$ in \Fig{fig:EEA_Combined_Fresh}(c) for $\tau_{23}\geq 0.3$\,ps, together with the same quantity deduced from the fit \Eq{eqn:ComplexFit} to the data shown in \Fig{fig:FWMDensityPowerFresh}. The linear dependence predicted by \Eq{eqn:EEAEquation} does not describe the data. Furthermore, the data for different $P$ do not overlap, as would be predicted by \Eq{eqn:EEAEquation}. Similar results were found for the aged sample using the data from \Fig{fig:FWMDensityPower}. We therefore conclude that EEA is not significant in our experiments.
\section{\label{Trion dynamics}Trion population dynamics}

\begin{figure}
	\centering
	\includegraphics[width=\columnwidth]{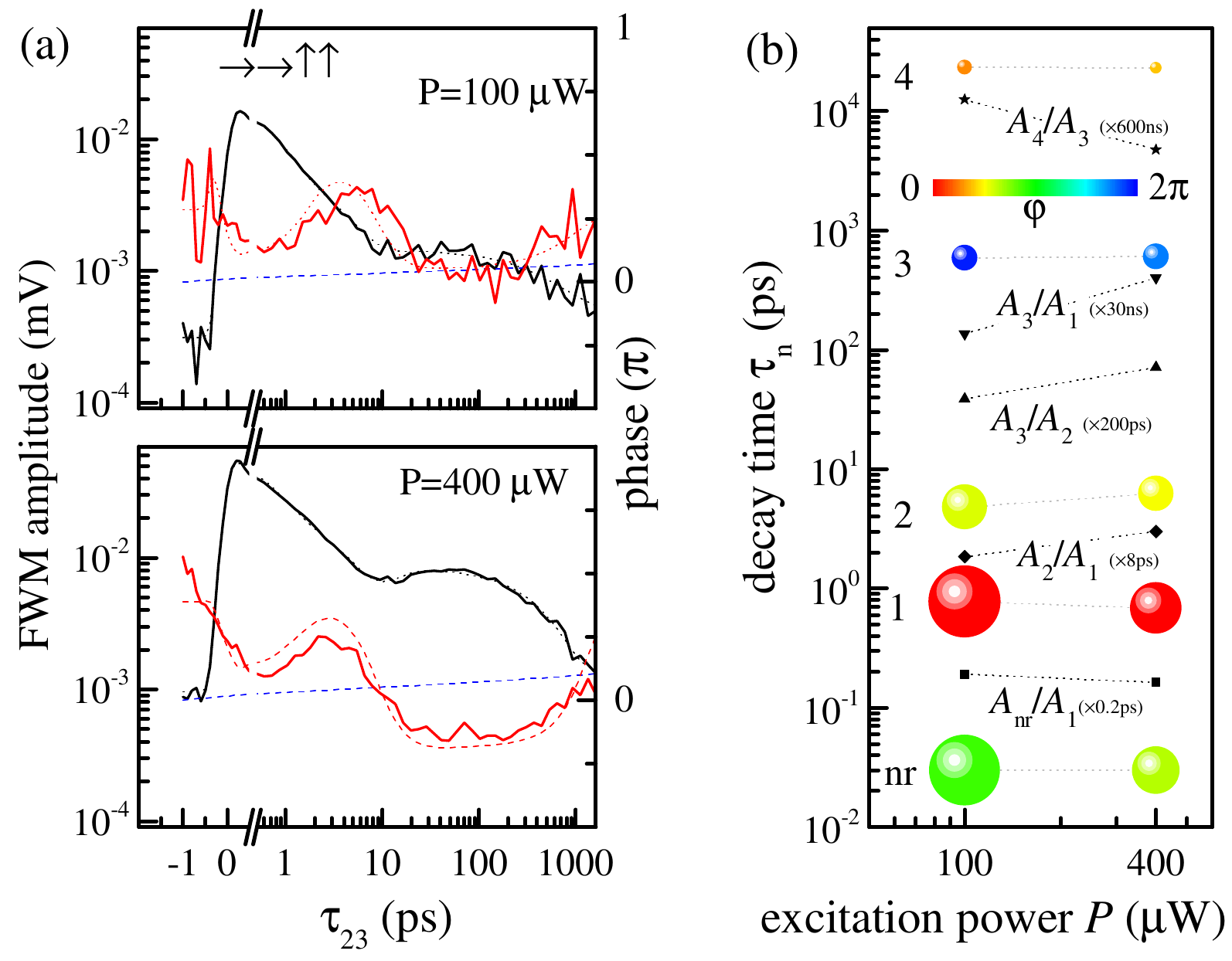}
	\caption{Density dynamics of the fresh sample for excitation resonant to $\XAT$ at T=77\,K. (a) FWM field amplitude and phase as function of the delay $\tau_{23}$ for two excitation powers $P$ as given and cross-linear polarization configuration $(\rightarrow, \rightarrow,\uparrow,\uparrow)$. The fitted phase drift (blue dashed line) has been subtracted from the data shown. (b) Results of a fit to (a), formatted as \Fig{fig:FitParaPower}. For absolute scaling $A_1=24\,\mu$V at $P=100\,\mu$W.}
	\label{fig:FWMDensityTrion}
\end{figure}

In order to investigate the trion $\XAT$ dynamics, we measured the population dynamics resonantly exciting and probing the $\XAT$ transition (see \Fig{fig:extinction}(b)) using cross-linearly polarized pumps at T=77\,K. The data are shown in \Fig{fig:FWMDensityTrion}(a) for two different powers. As in \Sec{sec:exciton}, we fitted the data using \Eq{eqn:ComplexFit} with four decay processes, and the resulting decay times, amplitudes and phases are shown in \Fig{fig:FWMDensityTrion}(b). We find an initial fast decay of $\tau_1\approx 1$\,ps similar to the exciton-resonant dynamics. The instantaneous response $A_{\rm nr}$ is stronger relative to $A_1$, and is inverse ($\varphi_{\rm nr}\approx \pi$) to $A_1$. The stronger relative weight is expected considering the weaker $\XAT$ absorption and about five times smaller $A_1$ for a given power. The second timescale $\tau_2\approx 5$\,ps is shorter than for the exciton-resonant dynamics where $\tau_2 \approx 20$\,ps, and has a larger relative amplitude compared to $A_1$. The later dynamics given by $\tau_3$ and $\tau_4$ instead are similar to what is found for the exciton-resonant excitation. We do not observe a significant variation of the timescales with power, in difference to what has been recently reported\,\cite{GaoPRB16} and interpreted as saturation of defect states. The amplitudes slightly decrease with increasing power. All the amplitude ratios seem to increase with the power, with the exception of $A_4/A_3$. The phase evolution is quite different from the exciton-resonant data - we find $\varphi_2\approx \pi/2$ while $\varphi_3\sim 2\pi$, equvalent to zero. $\varphi_4$ is slightly less then $\pi/2$. The similarity of the $\tau_3$ and $\tau_4$ values to the exciton-resonant results indicates that after about 100\,ps, the initially excited $\XAT$ are thermalized with $\XA$, and thus the memory of the initially excited transition is lost.

Concerning the initial dynamics, we note that exciton to trion conversion has been recently investigated\,\cite{SinghPRB16}, showing a conversion within 1-2\,ps at T=13\,K. At this temperature, the trion with a binding energy of 30\,meV is not thermally excited, so that the time measured is the trion formation time. We emphasize that the rate of the corresponding process exciton plus electron gives trion plus phonon is expected to be proportional to the electron density, and the latter was not given in\,\cite{SinghPRB16}. The reported timescale should therefore be taken only as a possible value, which can vary strongly depending on the electron density.     

In our experiment we initally excite $\XAT$, and we are therefore measuring as initial process the thermal excitation of the trion within its dispersion, and its ionization into a free electron and an exciton.
The radiative recombination of trions is expected to be much slower than the slow bright excitons, due to the momentum distribution of the remaining electron after the decay\,\cite{EsserPRB00}. We therefore attribute the first timescale $\tau_1$ to trion thermalization within its dispersion, as trion ionization is not expected to be efficient at $T=77$\,K, due to the large trion binding energy of 30\,meV compared to the thermal energy. The second timescale $\tau_2$ should then be attributed to trion ionization, a process which was recently observed in photoluminescence\,\cite{JonesNPhy15}. After ionization, the dynamics are expected to be similar to the case of exciton excitation, and in fact we see similar timescales. The phases are somewhat different due to the difference in energy of the probed optical response. 

\section{\label{Conclusions}Conclusions}

In conclusion, we have measured the exciton dephasing and density dynamics in SL-$\text{MoSe}_2$ using resonant excitation, which is avoiding the excitation of free electron-hole pairs. We determined at $T=77\,$K a dephasing time of 0.36\,ps for the exciton and 0.27\,ps for the trion. The measured density density dynamics from 100\,fs to 10\,ns shows 4 distinct processes which we interpreted in terms of the intrinsic band structure as radiative decay, scattering within the exciton dispersion, and scattering between bright direct and dark indirect excitons. Exciton-exciton annihilation was excluded as relevant process in the dynamics.  

\acknowledgments This work was partially funded by the EPSRC under grant EP/M020479/1, EP/M012727/1, and by the Graphene Flagship under grant agreement 696656. W.L. thanks J. Kasprzak for helpful discussions. FW acknowledges support from the Royal Academy of Engineering.

\bibliography{MyReferences,langsrv}

\begin{thebibliography}{48}%
\makeatletter
\providecommand \@ifxundefined [1]{%
 \@ifx{#1\undefined}
}%
\providecommand \@ifnum [1]{%
 \ifnum #1\expandafter \@firstoftwo
 \else \expandafter \@secondoftwo
 \fi
}%
\providecommand \@ifx [1]{%
 \ifx #1\expandafter \@firstoftwo
 \else \expandafter \@secondoftwo
 \fi
}%
\providecommand \natexlab [1]{#1}%
\providecommand \enquote  [1]{``#1''}%
\providecommand \bibnamefont  [1]{#1}%
\providecommand \bibfnamefont [1]{#1}%
\providecommand \citenamefont [1]{#1}%
\providecommand \href@noop [0]{\@secondoftwo}%
\providecommand \href [0]{\begingroup \@sanitize@url \@href}%
\providecommand \@href[1]{\@@startlink{#1}\@@href}%
\providecommand \@@href[1]{\endgroup#1\@@endlink}%
\providecommand \@sanitize@url [0]{\catcode `\\12\catcode `\$12\catcode
  `\&12\catcode `\#12\catcode `\^12\catcode `\_12\catcode `\%12\relax}%
\providecommand \@@startlink[1]{}%
\providecommand \@@endlink[0]{}%
\providecommand \url  [0]{\begingroup\@sanitize@url \@url }%
\providecommand \@url [1]{\endgroup\@href {#1}{\urlprefix }}%
\providecommand \urlprefix  [0]{URL }%
\providecommand \Eprint [0]{\href }%
\providecommand \doibase [0]{http://dx.doi.org/}%
\providecommand \selectlanguage [0]{\@gobble}%
\providecommand \bibinfo  [0]{\@secondoftwo}%
\providecommand \bibfield  [0]{\@secondoftwo}%
\providecommand \translation [1]{[#1]}%
\providecommand \BibitemOpen [0]{}%
\providecommand \bibitemStop [0]{}%
\providecommand \bibitemNoStop [0]{.\EOS\space}%
\providecommand \EOS [0]{\spacefactor3000\relax}%
\providecommand \BibitemShut  [1]{\csname bibitem#1\endcsname}%
\let\auto@bib@innerbib\@empty
\bibitem [{\citenamefont {Jakubczyk}\ \emph {et~al.}(2016)\citenamefont
  {Jakubczyk}, \citenamefont {Delmonte}, \citenamefont {Koperski},
  \citenamefont {Nogajewski}, \citenamefont {Faugeras}, \citenamefont
  {Langbein}, \citenamefont {Potemski},\ and\ \citenamefont
  {Kasprzak}}]{JakubczykNL16}%
  \BibitemOpen
  \bibfield  {author} {\bibinfo {author} {\bibfnamefont {T.}~\bibnamefont
  {Jakubczyk}}, \bibinfo {author} {\bibfnamefont {V.}~\bibnamefont {Delmonte}},
  \bibinfo {author} {\bibfnamefont {M.}~\bibnamefont {Koperski}}, \bibinfo
  {author} {\bibfnamefont {K.}~\bibnamefont {Nogajewski}}, \bibinfo {author}
  {\bibfnamefont {C.}~\bibnamefont {Faugeras}}, \bibinfo {author}
  {\bibfnamefont {W.}~\bibnamefont {Langbein}}, \bibinfo {author}
  {\bibfnamefont {M.}~\bibnamefont {Potemski}}, \ and\ \bibinfo {author}
  {\bibfnamefont {J.}~\bibnamefont {Kasprzak}},\ }\href@noop {} {\bibfield
  {journal} {\bibinfo  {journal} {Nano Lett.}\ }\textbf {\bibinfo {volume}
  {10}},\ \bibinfo {pages} {1021} (\bibinfo {year} {2016})}\BibitemShut
  {NoStop}%
\bibitem [{\citenamefont {Mak}\ \emph {et~al.}(2010)\citenamefont {Mak},
  \citenamefont {Lee}, \citenamefont {Hone}, \citenamefont {Shan},\ and\
  \citenamefont {Heinz}}]{MakPRL10}%
  \BibitemOpen
  \bibfield  {author} {\bibinfo {author} {\bibfnamefont {K.~F.}\ \bibnamefont
  {Mak}}, \bibinfo {author} {\bibfnamefont {C.}~\bibnamefont {Lee}}, \bibinfo
  {author} {\bibfnamefont {J.}~\bibnamefont {Hone}}, \bibinfo {author}
  {\bibfnamefont {J.}~\bibnamefont {Shan}}, \ and\ \bibinfo {author}
  {\bibfnamefont {T.~F.}\ \bibnamefont {Heinz}},\ }\href@noop {} {\bibfield
  {journal} {\bibinfo  {journal} {Phys. Rev. Lett.}\ }\textbf {\bibinfo
  {volume} {105}},\ \bibinfo {pages} {136805} (\bibinfo {year}
  {2010})}\BibitemShut {NoStop}%
\bibitem [{\citenamefont {Splendiani}\ \emph {et~al.}(2010)\citenamefont
  {Splendiani}, \citenamefont {Sun}, \citenamefont {Zhang}, \citenamefont {Li},
  \citenamefont {Kim}, \citenamefont {Chim}, \citenamefont {Galli},\ and\
  \citenamefont {Wang}}]{SplendianiNL10}%
  \BibitemOpen
  \bibfield  {author} {\bibinfo {author} {\bibfnamefont {A.}~\bibnamefont
  {Splendiani}}, \bibinfo {author} {\bibfnamefont {L.}~\bibnamefont {Sun}},
  \bibinfo {author} {\bibfnamefont {Y.}~\bibnamefont {Zhang}}, \bibinfo
  {author} {\bibfnamefont {T.}~\bibnamefont {Li}}, \bibinfo {author}
  {\bibfnamefont {J.}~\bibnamefont {Kim}}, \bibinfo {author} {\bibfnamefont
  {C.-Y.}\ \bibnamefont {Chim}}, \bibinfo {author} {\bibfnamefont
  {G.}~\bibnamefont {Galli}}, \ and\ \bibinfo {author} {\bibfnamefont
  {F.}~\bibnamefont {Wang}},\ }\href@noop {} {\bibfield  {journal} {\bibinfo
  {journal} {Nano Lett.}\ }\textbf {\bibinfo {volume} {10}},\ \bibinfo {pages}
  {1271} (\bibinfo {year} {2010})}\BibitemShut {NoStop}%
\bibitem [{\citenamefont {Xiao}\ \emph {et~al.}(2012)\citenamefont {Xiao},
  \citenamefont {Liu}, \citenamefont {Feng}, \citenamefont {Xu},\ and\
  \citenamefont {Yao}}]{XiaoPRL12}%
  \BibitemOpen
  \bibfield  {author} {\bibinfo {author} {\bibfnamefont {D.}~\bibnamefont
  {Xiao}}, \bibinfo {author} {\bibfnamefont {G.-B.}\ \bibnamefont {Liu}},
  \bibinfo {author} {\bibfnamefont {W.}~\bibnamefont {Feng}}, \bibinfo {author}
  {\bibfnamefont {X.}~\bibnamefont {Xu}}, \ and\ \bibinfo {author}
  {\bibfnamefont {W.}~\bibnamefont {Yao}},\ }\href@noop {} {\bibfield
  {journal} {\bibinfo  {journal} {Phys. Rev. Lett.}\ }\textbf {\bibinfo
  {volume} {108}},\ \bibinfo {pages} {196802} (\bibinfo {year}
  {2012})}\BibitemShut {NoStop}%
\bibitem [{\citenamefont {Mak}\ \emph {et~al.}(2012)\citenamefont {Mak},
  \citenamefont {He}, \citenamefont {Shan},\ and\ \citenamefont
  {Heinz}}]{MakNNa12}%
  \BibitemOpen
  \bibfield  {author} {\bibinfo {author} {\bibfnamefont {K.~F.}\ \bibnamefont
  {Mak}}, \bibinfo {author} {\bibfnamefont {K.}~\bibnamefont {He}}, \bibinfo
  {author} {\bibfnamefont {J.}~\bibnamefont {Shan}}, \ and\ \bibinfo {author}
  {\bibfnamefont {T.~F.}\ \bibnamefont {Heinz}},\ }\href@noop {} {\bibfield
  {journal} {\bibinfo  {journal} {Nat. Nanotech.}\ }\textbf {\bibinfo {volume}
  {7}},\ \bibinfo {pages} {494} (\bibinfo {year} {2012})}\BibitemShut {NoStop}%
\bibitem [{\citenamefont {Ugeda}\ \emph {et~al.}(2014)\citenamefont {Ugeda},
  \citenamefont {Bradley}, \citenamefont {Shi}, \citenamefont {da~Jornada},
  \citenamefont {Zhang}, \citenamefont {Qiu}, \citenamefont {Ruan},
  \citenamefont {Mo}, \citenamefont {Hussain}, \citenamefont {Shen},
  \citenamefont {FengWang}, \citenamefont {Louie},\ and\ \citenamefont
  {Crommie}}]{UgedaNMa14}%
  \BibitemOpen
  \bibfield  {author} {\bibinfo {author} {\bibfnamefont {M.~M.}\ \bibnamefont
  {Ugeda}}, \bibinfo {author} {\bibfnamefont {A.~J.}\ \bibnamefont {Bradley}},
  \bibinfo {author} {\bibfnamefont {S.-F.}\ \bibnamefont {Shi}}, \bibinfo
  {author} {\bibfnamefont {F.~H.}\ \bibnamefont {da~Jornada}}, \bibinfo
  {author} {\bibfnamefont {Y.}~\bibnamefont {Zhang}}, \bibinfo {author}
  {\bibfnamefont {D.~Y.}\ \bibnamefont {Qiu}}, \bibinfo {author} {\bibfnamefont
  {W.}~\bibnamefont {Ruan}}, \bibinfo {author} {\bibfnamefont {S.-K.}\
  \bibnamefont {Mo}}, \bibinfo {author} {\bibfnamefont {Z.}~\bibnamefont
  {Hussain}}, \bibinfo {author} {\bibfnamefont {Z.-X.}\ \bibnamefont {Shen}},
  \bibinfo {author} {\bibnamefont {FengWang}}, \bibinfo {author} {\bibfnamefont
  {S.~G.}\ \bibnamefont {Louie}}, \ and\ \bibinfo {author} {\bibfnamefont
  {M.~F.}\ \bibnamefont {Crommie}},\ }\href@noop {} {\bibfield  {journal}
  {\bibinfo  {journal} {Nat. Mater.}\ }\textbf {\bibinfo {volume} {13}},\
  \bibinfo {pages} {1091} (\bibinfo {year} {2014})}\BibitemShut {NoStop}%
\bibitem [{\citenamefont {Mouri}\ \emph {et~al.}(2014)\citenamefont {Mouri},
  \citenamefont {Miyauch}, \citenamefont {Toh}, \citenamefont {Zhao},
  \citenamefont {Eda},\ and\ \citenamefont {Matsuda}}]{MouriPRB14}%
  \BibitemOpen
  \bibfield  {author} {\bibinfo {author} {\bibfnamefont {S.}~\bibnamefont
  {Mouri}}, \bibinfo {author} {\bibfnamefont {Y.}~\bibnamefont {Miyauch}},
  \bibinfo {author} {\bibfnamefont {M.}~\bibnamefont {Toh}}, \bibinfo {author}
  {\bibfnamefont {W.}~\bibnamefont {Zhao}}, \bibinfo {author} {\bibfnamefont
  {G.}~\bibnamefont {Eda}}, \ and\ \bibinfo {author} {\bibfnamefont
  {K.}~\bibnamefont {Matsuda}},\ }\href@noop {} {\bibfield  {journal} {\bibinfo
   {journal} {Phys. Rev. B}\ }\textbf {\bibinfo {volume} {90}},\ \bibinfo
  {pages} {155449} (\bibinfo {year} {2014})}\BibitemShut {NoStop}%
\bibitem [{\citenamefont {Sun}\ \emph {et~al.}(2014)\citenamefont {Sun},
  \citenamefont {Rao}, \citenamefont {Reider}, \citenamefont {Chen},
  \citenamefont {You}, \citenamefont {Br\'{e}zin}, \citenamefont
  {Harutyunyan},\ and\ \citenamefont {Heinz}}]{SunNL14}%
  \BibitemOpen
  \bibfield  {author} {\bibinfo {author} {\bibfnamefont {D.}~\bibnamefont
  {Sun}}, \bibinfo {author} {\bibfnamefont {Y.}~\bibnamefont {Rao}}, \bibinfo
  {author} {\bibfnamefont {G.~A.}\ \bibnamefont {Reider}}, \bibinfo {author}
  {\bibfnamefont {G.}~\bibnamefont {Chen}}, \bibinfo {author} {\bibfnamefont
  {Y.}~\bibnamefont {You}}, \bibinfo {author} {\bibfnamefont {L.}~\bibnamefont
  {Br\'{e}zin}}, \bibinfo {author} {\bibfnamefont {A.~R.}\ \bibnamefont
  {Harutyunyan}}, \ and\ \bibinfo {author} {\bibfnamefont {T.~F.}\ \bibnamefont
  {Heinz}},\ }\href@noop {} {\bibfield  {journal} {\bibinfo  {journal} {Nano
  Lett.}\ }\textbf {\bibinfo {volume} {14}},\ \bibinfo {pages} {5625} (\bibinfo
  {year} {2014})}\BibitemShut {NoStop}%
\bibitem [{\citenamefont {Kumar}\ \emph {et~al.}(2014)\citenamefont {Kumar},
  \citenamefont {Cui}, \citenamefont {Ceballos}, \citenamefont {He},
  \citenamefont {Wang},\ and\ \citenamefont {Zhao}}]{KumarPRB14}%
  \BibitemOpen
  \bibfield  {author} {\bibinfo {author} {\bibfnamefont {N.}~\bibnamefont
  {Kumar}}, \bibinfo {author} {\bibfnamefont {Q.}~\bibnamefont {Cui}}, \bibinfo
  {author} {\bibfnamefont {F.}~\bibnamefont {Ceballos}}, \bibinfo {author}
  {\bibfnamefont {D.}~\bibnamefont {He}}, \bibinfo {author} {\bibfnamefont
  {Y.}~\bibnamefont {Wang}}, \ and\ \bibinfo {author} {\bibfnamefont
  {H.}~\bibnamefont {Zhao}},\ }\href@noop {} {\bibfield  {journal} {\bibinfo
  {journal} {Phys. Rev. B}\ }\textbf {\bibinfo {volume} {89}},\ \bibinfo
  {pages} {125427} (\bibinfo {year} {2014})}\BibitemShut {NoStop}%
\bibitem [{\citenamefont {Zhu}\ \emph {et~al.}(2011)\citenamefont {Zhu},
  \citenamefont {Cheng},\ and\ \citenamefont
  {Schwingenschl\"{o}gl}}]{ZhuPRB11}%
  \BibitemOpen
  \bibfield  {author} {\bibinfo {author} {\bibfnamefont {Z.~Y.}\ \bibnamefont
  {Zhu}}, \bibinfo {author} {\bibfnamefont {Y.~C.}\ \bibnamefont {Cheng}}, \
  and\ \bibinfo {author} {\bibfnamefont {U.}~\bibnamefont
  {Schwingenschl\"{o}gl}},\ }\href@noop {} {\bibfield  {journal} {\bibinfo
  {journal} {Phys. Rev. B}\ }\textbf {\bibinfo {volume} {84}},\ \bibinfo
  {pages} {153402} (\bibinfo {year} {2011})}\BibitemShut {NoStop}%
\bibitem [{\citenamefont {Ko\'{s}mider}\ \emph {et~al.}(2013)\citenamefont
  {Ko\'{s}mider}, \citenamefont {Gonz\'{a}lez},\ and\ \citenamefont
  {Fern\'{a}ndez-Rossier}}]{KosmiderPRB13}%
  \BibitemOpen
  \bibfield  {author} {\bibinfo {author} {\bibfnamefont {K.}~\bibnamefont
  {Ko\'{s}mider}}, \bibinfo {author} {\bibfnamefont {J.~W.}\ \bibnamefont
  {Gonz\'{a}lez}}, \ and\ \bibinfo {author} {\bibfnamefont {J.}~\bibnamefont
  {Fern\'{a}ndez-Rossier}},\ }\href@noop {} {\bibfield  {journal} {\bibinfo
  {journal} {Phys. Rev. B}\ }\textbf {\bibinfo {volume} {88}},\ \bibinfo
  {pages} {245436} (\bibinfo {year} {2013})}\BibitemShut {NoStop}%
\bibitem [{\citenamefont {Zhang}\ \emph
  {et~al.}(2015{\natexlab{a}})\citenamefont {Zhang}, \citenamefont {You},
  \citenamefont {Zhao},\ and\ \citenamefont {Heinz}}]{ZhangPRL15}%
  \BibitemOpen
  \bibfield  {author} {\bibinfo {author} {\bibfnamefont {X.-X.}\ \bibnamefont
  {Zhang}}, \bibinfo {author} {\bibfnamefont {Y.}~\bibnamefont {You}}, \bibinfo
  {author} {\bibfnamefont {S.~Y.~F.}\ \bibnamefont {Zhao}}, \ and\ \bibinfo
  {author} {\bibfnamefont {T.~F.}\ \bibnamefont {Heinz}},\ }\href@noop {}
  {\bibfield  {journal} {\bibinfo  {journal} {Phys. Rev. Lett.}\ }\textbf
  {\bibinfo {volume} {115}},\ \bibinfo {pages} {257403} (\bibinfo {year}
  {2015}{\natexlab{a}})}\BibitemShut {NoStop}%
\bibitem [{\citenamefont {Liu}\ \emph {et~al.}(2013)\citenamefont {Liu},
  \citenamefont {Shan}, \citenamefont {Yao}, \citenamefont {Yao},\ and\
  \citenamefont {Xiao}}]{LiuPRB13}%
  \BibitemOpen
  \bibfield  {author} {\bibinfo {author} {\bibfnamefont {G.-B.}\ \bibnamefont
  {Liu}}, \bibinfo {author} {\bibfnamefont {W.-Y.}\ \bibnamefont {Shan}},
  \bibinfo {author} {\bibfnamefont {Y.}~\bibnamefont {Yao}}, \bibinfo {author}
  {\bibfnamefont {W.}~\bibnamefont {Yao}}, \ and\ \bibinfo {author}
  {\bibfnamefont {D.}~\bibnamefont {Xiao}},\ }\href@noop {} {\bibfield
  {journal} {\bibinfo  {journal} {Phys. Rev. B}\ }\textbf {\bibinfo {volume}
  {88}},\ \bibinfo {pages} {085433} (\bibinfo {year} {2013})}\BibitemShut
  {NoStop}%
\bibitem [{\citenamefont {Dery}\ and\ \citenamefont {Song}(2015)}]{DeryPRB15}%
  \BibitemOpen
  \bibfield  {author} {\bibinfo {author} {\bibfnamefont {H.}~\bibnamefont
  {Dery}}\ and\ \bibinfo {author} {\bibfnamefont {Y.}~\bibnamefont {Song}},\
  }\href@noop {} {\bibfield  {journal} {\bibinfo  {journal} {Phys. Rev. B}\
  }\textbf {\bibinfo {volume} {92}},\ \bibinfo {pages} {125431} (\bibinfo
  {year} {2015})}\BibitemShut {NoStop}%
\bibitem [{\citenamefont {Zhang}\ \emph
  {et~al.}(2015{\natexlab{b}})\citenamefont {Zhang}, \citenamefont {Fan},
  \citenamefont {Li}, \citenamefont {Ji}, \citenamefont {Zhao}, \citenamefont
  {Xia}, \citenamefont {Yan}, \citenamefont {Zhang}, \citenamefont {Zhang},
  \citenamefont {Wang},\ and\ \citenamefont {Zhang}}]{ZhangARX15}%
  \BibitemOpen
  \bibfield  {author} {\bibinfo {author} {\bibfnamefont {A.}~\bibnamefont
  {Zhang}}, \bibinfo {author} {\bibfnamefont {J.}~\bibnamefont {Fan}}, \bibinfo
  {author} {\bibfnamefont {Y.}~\bibnamefont {Li}}, \bibinfo {author}
  {\bibfnamefont {J.}~\bibnamefont {Ji}}, \bibinfo {author} {\bibfnamefont
  {G.}~\bibnamefont {Zhao}}, \bibinfo {author} {\bibfnamefont {T.}~\bibnamefont
  {Xia}}, \bibinfo {author} {\bibfnamefont {T.}~\bibnamefont {Yan}}, \bibinfo
  {author} {\bibfnamefont {X.}~\bibnamefont {Zhang}}, \bibinfo {author}
  {\bibfnamefont {W.}~\bibnamefont {Zhang}}, \bibinfo {author} {\bibfnamefont
  {X.}~\bibnamefont {Wang}}, \ and\ \bibinfo {author} {\bibfnamefont
  {Q.}~\bibnamefont {Zhang}},\ }\href@noop {} {\bibfield  {journal} {\bibinfo
  {journal} {arXiv:1503.08631}\ } (\bibinfo {year}
  {2015}{\natexlab{b}})}\BibitemShut {NoStop}%
\bibitem [{\citenamefont {Dufferwiel}\ \emph {et~al.}(2016)\citenamefont
  {Dufferwiel}, \citenamefont {Lyons}, \citenamefont {Solnyshkov},
  \citenamefont {Trichet}, \citenamefont {Withers}, \citenamefont {Schwarz},
  \citenamefont {Malpuech}, \citenamefont {Smith}, \citenamefont {Novoselov},
  \citenamefont {Skolnick}, \citenamefont {Krizhanovskii},\ and\ \citenamefont
  {Tartakovskii}}]{DufferwielARX16}%
  \BibitemOpen
  \bibfield  {author} {\bibinfo {author} {\bibfnamefont {S.}~\bibnamefont
  {Dufferwiel}}, \bibinfo {author} {\bibfnamefont {T.~P.}\ \bibnamefont
  {Lyons}}, \bibinfo {author} {\bibfnamefont {D.~D.}\ \bibnamefont
  {Solnyshkov}}, \bibinfo {author} {\bibfnamefont {A.~A.~P.}\ \bibnamefont
  {Trichet}}, \bibinfo {author} {\bibfnamefont {F.}~\bibnamefont {Withers}},
  \bibinfo {author} {\bibfnamefont {S.}~\bibnamefont {Schwarz}}, \bibinfo
  {author} {\bibfnamefont {G.}~\bibnamefont {Malpuech}}, \bibinfo {author}
  {\bibfnamefont {J.~M.}\ \bibnamefont {Smith}}, \bibinfo {author}
  {\bibfnamefont {K.~S.}\ \bibnamefont {Novoselov}}, \bibinfo {author}
  {\bibfnamefont {M.~S.}\ \bibnamefont {Skolnick}}, \bibinfo {author}
  {\bibfnamefont {D.~N.}\ \bibnamefont {Krizhanovskii}}, \ and\ \bibinfo
  {author} {\bibfnamefont {A.~I.}\ \bibnamefont {Tartakovskii}},\ }\href@noop
  {} {\bibfield  {journal} {\bibinfo  {journal} {arXiv:1612.05073}\ } (\bibinfo
  {year} {2016})}\BibitemShut {NoStop}%
\bibitem [{\citenamefont {Robert}\ \emph {et~al.}(2016)\citenamefont {Robert},
  \citenamefont {Lagarde}, \citenamefont {Cadiz}, \citenamefont {Wang},
  \citenamefont {Lassagne}, \citenamefont {Amand}, \citenamefont {Balocchi},
  \citenamefont {Renucci}, \citenamefont {Tongay}, \citenamefont {Urbaszek},\
  and\ \citenamefont {Marie}}]{RobertARX16}%
  \BibitemOpen
  \bibfield  {author} {\bibinfo {author} {\bibfnamefont {C.}~\bibnamefont
  {Robert}}, \bibinfo {author} {\bibfnamefont {D.}~\bibnamefont {Lagarde}},
  \bibinfo {author} {\bibfnamefont {F.}~\bibnamefont {Cadiz}}, \bibinfo
  {author} {\bibfnamefont {G.}~\bibnamefont {Wang}}, \bibinfo {author}
  {\bibfnamefont {B.}~\bibnamefont {Lassagne}}, \bibinfo {author}
  {\bibfnamefont {T.}~\bibnamefont {Amand}}, \bibinfo {author} {\bibfnamefont
  {A.}~\bibnamefont {Balocchi}}, \bibinfo {author} {\bibfnamefont
  {P.}~\bibnamefont {Renucci}}, \bibinfo {author} {\bibfnamefont
  {S.}~\bibnamefont {Tongay}}, \bibinfo {author} {\bibfnamefont
  {B.}~\bibnamefont {Urbaszek}}, \ and\ \bibinfo {author} {\bibfnamefont
  {X.}~\bibnamefont {Marie}},\ }\href@noop {} {\bibfield  {journal} {\bibinfo
  {journal} {Phys. Rev. B}\ }\textbf {\bibinfo {volume} {93}},\ \bibinfo
  {pages} {205423} (\bibinfo {year} {2016})}\BibitemShut {NoStop}%
\bibitem [{\citenamefont {Selig}\ \emph {et~al.}(2016)\citenamefont {Selig},
  \citenamefont {Berghäuser}, \citenamefont {Raja}, \citenamefont {Nagler},
  \citenamefont {Schüller}, \citenamefont {Heinz}, \citenamefont {Korn},
  \citenamefont {Chernikov}, \citenamefont {Malic},\ and\ \citenamefont
  {Knorr}}]{SeligNC16}%
  \BibitemOpen
  \bibfield  {author} {\bibinfo {author} {\bibfnamefont {M.}~\bibnamefont
  {Selig}}, \bibinfo {author} {\bibfnamefont {G.}~\bibnamefont {Berghäuser}},
  \bibinfo {author} {\bibfnamefont {A.}~\bibnamefont {Raja}}, \bibinfo {author}
  {\bibfnamefont {P.}~\bibnamefont {Nagler}}, \bibinfo {author} {\bibfnamefont
  {C.}~\bibnamefont {Schüller}}, \bibinfo {author} {\bibfnamefont {T.~F.}\
  \bibnamefont {Heinz}}, \bibinfo {author} {\bibfnamefont {T.}~\bibnamefont
  {Korn}}, \bibinfo {author} {\bibfnamefont {A.}~\bibnamefont {Chernikov}},
  \bibinfo {author} {\bibfnamefont {E.}~\bibnamefont {Malic}}, \ and\ \bibinfo
  {author} {\bibfnamefont {A.}~\bibnamefont {Knorr}},\ }\href {\doibase
  10.1038/ncomms13279} {\bibfield  {journal} {\bibinfo  {journal} {Nature
  Communications}\ }\textbf {\bibinfo {volume} {7}},\ \bibinfo {pages} {13279}
  (\bibinfo {year} {2016})}\BibitemShut {NoStop}%
\bibitem [{\citenamefont {Moody}\ \emph {et~al.}(2016)\citenamefont {Moody},
  \citenamefont {Schaibley},\ and\ \citenamefont {Xu}}]{MoodyJOSAB16}%
  \BibitemOpen
  \bibfield  {author} {\bibinfo {author} {\bibfnamefont {G.}~\bibnamefont
  {Moody}}, \bibinfo {author} {\bibfnamefont {J.}~\bibnamefont {Schaibley}}, \
  and\ \bibinfo {author} {\bibfnamefont {X.}~\bibnamefont {Xu}},\ }\href@noop
  {} {\bibfield  {journal} {\bibinfo  {journal} {J. Opt. Soc. Am. B}\ }\textbf
  {\bibinfo {volume} {33}},\ \bibinfo {pages} {C39} (\bibinfo {year}
  {2016})}\BibitemShut {NoStop}%
\bibitem [{\citenamefont {Yu}\ \emph {et~al.}(2014)\citenamefont {Yu},
  \citenamefont {Liu}, \citenamefont {Gong}, \citenamefont {Xu},\ and\
  \citenamefont {Yao}}]{YuNC14}%
  \BibitemOpen
  \bibfield  {author} {\bibinfo {author} {\bibfnamefont {H.}~\bibnamefont
  {Yu}}, \bibinfo {author} {\bibfnamefont {G.-B.}\ \bibnamefont {Liu}},
  \bibinfo {author} {\bibfnamefont {P.}~\bibnamefont {Gong}}, \bibinfo {author}
  {\bibfnamefont {X.}~\bibnamefont {Xu}}, \ and\ \bibinfo {author}
  {\bibfnamefont {W.}~\bibnamefont {Yao}},\ }\href@noop {} {\bibfield
  {journal} {\bibinfo  {journal} {Nat. Commun.}\ }\textbf {\bibinfo {volume}
  {5}},\ \bibinfo {pages} {3876} (\bibinfo {year} {2014})}\BibitemShut
  {NoStop}%
\bibitem [{\citenamefont {Ross}\ \emph {et~al.}(2013)\citenamefont {Ross},
  \citenamefont {Wu}, \citenamefont {Yu}, \citenamefont {Ghimire},
  \citenamefont {Jones}, \citenamefont {Aivazian}, \citenamefont {Yan},
  \citenamefont {Mandrus}, \citenamefont {Xiao}, \citenamefont {Yao},\ and\
  \citenamefont {Xu}}]{RossNC13}%
  \BibitemOpen
  \bibfield  {author} {\bibinfo {author} {\bibfnamefont {J.~S.}\ \bibnamefont
  {Ross}}, \bibinfo {author} {\bibfnamefont {S.}~\bibnamefont {Wu}}, \bibinfo
  {author} {\bibfnamefont {H.}~\bibnamefont {Yu}}, \bibinfo {author}
  {\bibfnamefont {N.~J.}\ \bibnamefont {Ghimire}}, \bibinfo {author}
  {\bibfnamefont {A.~M.}\ \bibnamefont {Jones}}, \bibinfo {author}
  {\bibfnamefont {G.}~\bibnamefont {Aivazian}}, \bibinfo {author}
  {\bibfnamefont {J.}~\bibnamefont {Yan}}, \bibinfo {author} {\bibfnamefont
  {D.~G.}\ \bibnamefont {Mandrus}}, \bibinfo {author} {\bibfnamefont
  {D.}~\bibnamefont {Xiao}}, \bibinfo {author} {\bibfnamefont {W.}~\bibnamefont
  {Yao}}, \ and\ \bibinfo {author} {\bibfnamefont {X.}~\bibnamefont {Xu}},\
  }\href@noop {} {\bibfield  {journal} {\bibinfo  {journal} {Nat. Commun.}\
  }\textbf {\bibinfo {volume} {4}},\ \bibinfo {pages} {1474} (\bibinfo {year}
  {2013})}\BibitemShut {NoStop}%
\bibitem [{\citenamefont {Esser}\ \emph {et~al.}(2000)\citenamefont {Esser},
  \citenamefont {Runge}, \citenamefont {Zimmermann},\ and\ \citenamefont
  {Langbein}}]{EsserPRB00}%
  \BibitemOpen
  \bibfield  {author} {\bibinfo {author} {\bibfnamefont {A.}~\bibnamefont
  {Esser}}, \bibinfo {author} {\bibfnamefont {E.}~\bibnamefont {Runge}},
  \bibinfo {author} {\bibfnamefont {R.}~\bibnamefont {Zimmermann}}, \ and\
  \bibinfo {author} {\bibfnamefont {W.}~\bibnamefont {Langbein}},\ }\href@noop
  {} {\bibfield  {journal} {\bibinfo  {journal} {Phys. Rev. B}\ }\textbf
  {\bibinfo {volume} {62}},\ \bibinfo {pages} {8232} (\bibinfo {year}
  {2000})}\BibitemShut {NoStop}%
\bibitem [{\citenamefont {Sidler}\ \emph {et~al.}(2017)\citenamefont {Sidler},
  \citenamefont {Back}, \citenamefont {Cotlet}, \citenamefont {Srivastava},
  \citenamefont {Fink}, \citenamefont {Kroner}, \citenamefont {Demler},\ and\
  \citenamefont {Imamoglu}}]{SidlerNPhy16}%
  \BibitemOpen
  \bibfield  {author} {\bibinfo {author} {\bibfnamefont {M.}~\bibnamefont
  {Sidler}}, \bibinfo {author} {\bibfnamefont {P.}~\bibnamefont {Back}},
  \bibinfo {author} {\bibfnamefont {O.}~\bibnamefont {Cotlet}}, \bibinfo
  {author} {\bibfnamefont {A.}~\bibnamefont {Srivastava}}, \bibinfo {author}
  {\bibfnamefont {T.}~\bibnamefont {Fink}}, \bibinfo {author} {\bibfnamefont
  {M.}~\bibnamefont {Kroner}}, \bibinfo {author} {\bibfnamefont
  {E.}~\bibnamefont {Demler}}, \ and\ \bibinfo {author} {\bibfnamefont
  {A.}~\bibnamefont {Imamoglu}},\ }\href {\doibase 10.1038/nphys3949}
  {\bibfield  {journal} {\bibinfo  {journal} {Nat. Phys.}\ }\textbf {\bibinfo
  {volume} {13}},\ \bibinfo {pages} {255} (\bibinfo {year} {2017})}\BibitemShut
  {NoStop}%
\bibitem [{\citenamefont {Horzum}\ \emph {et~al.}(2013)\citenamefont {Horzum},
  \citenamefont {Sahin}, \citenamefont {Cahangirov}, \citenamefont {Cudazzo},
  \citenamefont {andT. Serin},\ and\ \citenamefont {Peeters}}]{HorzumPRB13}%
  \BibitemOpen
  \bibfield  {author} {\bibinfo {author} {\bibfnamefont {S.}~\bibnamefont
  {Horzum}}, \bibinfo {author} {\bibfnamefont {H.}~\bibnamefont {Sahin}},
  \bibinfo {author} {\bibfnamefont {S.}~\bibnamefont {Cahangirov}}, \bibinfo
  {author} {\bibfnamefont {P.}~\bibnamefont {Cudazzo}}, \bibinfo {author}
  {\bibfnamefont {A.~R.}\ \bibnamefont {andT. Serin}}, \ and\ \bibinfo {author}
  {\bibfnamefont {F.~M.}\ \bibnamefont {Peeters}},\ }\href@noop {} {\bibfield
  {journal} {\bibinfo  {journal} {Phys. Rev. B}\ }\textbf {\bibinfo {volume}
  {87}},\ \bibinfo {pages} {125415} (\bibinfo {year} {2013})}\BibitemShut
  {NoStop}%
\bibitem [{\citenamefont {Singh}\ \emph {et~al.}(2014)\citenamefont {Singh},
  \citenamefont {Moody}, \citenamefont {Wu}, \citenamefont {Wu}, \citenamefont
  {Ghimire}, \citenamefont {Yan}, \citenamefont {Mandrus}, \citenamefont {Xu},\
  and\ \citenamefont {Li}}]{SinghPRL14}%
  \BibitemOpen
  \bibfield  {author} {\bibinfo {author} {\bibfnamefont {A.}~\bibnamefont
  {Singh}}, \bibinfo {author} {\bibfnamefont {G.}~\bibnamefont {Moody}},
  \bibinfo {author} {\bibfnamefont {S.}~\bibnamefont {Wu}}, \bibinfo {author}
  {\bibfnamefont {Y.}~\bibnamefont {Wu}}, \bibinfo {author} {\bibfnamefont
  {N.~J.}\ \bibnamefont {Ghimire}}, \bibinfo {author} {\bibfnamefont
  {J.}~\bibnamefont {Yan}}, \bibinfo {author} {\bibfnamefont {D.~G.}\
  \bibnamefont {Mandrus}}, \bibinfo {author} {\bibfnamefont {X.}~\bibnamefont
  {Xu}}, \ and\ \bibinfo {author} {\bibfnamefont {X.}~\bibnamefont {Li}},\
  }\href@noop {} {\bibfield  {journal} {\bibinfo  {journal} {Phys. Rev. Lett.}\
  }\textbf {\bibinfo {volume} {112}},\ \bibinfo {pages} {216804} (\bibinfo
  {year} {2014})}\BibitemShut {NoStop}%
\bibitem [{\citenamefont {Wang}\ \emph
  {et~al.}(2015{\natexlab{a}})\citenamefont {Wang}, \citenamefont {Palleau},
  \citenamefont {Amand}, \citenamefont {Tongay}, \citenamefont {Marie},\ and\
  \citenamefont {Urbaszek}}]{WangAPL15}%
  \BibitemOpen
  \bibfield  {author} {\bibinfo {author} {\bibfnamefont {G.}~\bibnamefont
  {Wang}}, \bibinfo {author} {\bibfnamefont {E.}~\bibnamefont {Palleau}},
  \bibinfo {author} {\bibfnamefont {T.}~\bibnamefont {Amand}}, \bibinfo
  {author} {\bibfnamefont {S.}~\bibnamefont {Tongay}}, \bibinfo {author}
  {\bibfnamefont {X.}~\bibnamefont {Marie}}, \ and\ \bibinfo {author}
  {\bibfnamefont {B.}~\bibnamefont {Urbaszek}},\ }\href@noop {} {\bibfield
  {journal} {\bibinfo  {journal} {App. Phys. Lett.}\ }\textbf {\bibinfo
  {volume} {106}},\ \bibinfo {pages} {112101} (\bibinfo {year}
  {2015}{\natexlab{a}})}\BibitemShut {NoStop}%
\bibitem [{\citenamefont {Gao}\ \emph {et~al.}(2016)\citenamefont {Gao},
  \citenamefont {Gong}, \citenamefont {Titze}, \citenamefont {Almeida},
  \citenamefont {Ajayan},\ and\ \citenamefont {Li}}]{GaoPRB16}%
  \BibitemOpen
  \bibfield  {author} {\bibinfo {author} {\bibfnamefont {F.}~\bibnamefont
  {Gao}}, \bibinfo {author} {\bibfnamefont {Y.}~\bibnamefont {Gong}}, \bibinfo
  {author} {\bibfnamefont {M.}~\bibnamefont {Titze}}, \bibinfo {author}
  {\bibfnamefont {R.}~\bibnamefont {Almeida}}, \bibinfo {author} {\bibfnamefont
  {P.~M.}\ \bibnamefont {Ajayan}}, \ and\ \bibinfo {author} {\bibfnamefont
  {H.}~\bibnamefont {Li}},\ }\href@noop {} {\bibfield  {journal} {\bibinfo
  {journal} {Phys. Rev. B}\ }\textbf {\bibinfo {volume} {94}},\ \bibinfo
  {pages} {245413} (\bibinfo {year} {2016})}\BibitemShut {NoStop}%
\bibitem [{\citenamefont {Borri}\ and\ \citenamefont
  {Langbein}(2009)}]{BorriBook09}%
  \BibitemOpen
  \bibfield  {author} {\bibinfo {author} {\bibfnamefont {P.}~\bibnamefont
  {Borri}}\ and\ \bibinfo {author} {\bibfnamefont {W.}~\bibnamefont
  {Langbein}},\ }in\ \href@noop {} {\emph {\bibinfo {booktitle} {Semiconductor
  Quantum Bits}}},\ \bibinfo {editor} {edited by\ \bibinfo {editor}
  {\bibfnamefont {O.}~\bibnamefont {Benson}}\ and\ \bibinfo {editor}
  {\bibfnamefont {F.}~\bibnamefont {Henneberger}}}\ (\bibinfo  {publisher}
  {World Scientific},\ \bibinfo {address} {Singapore},\ \bibinfo {year}
  {2009})\BibitemShut {NoStop}%
\bibitem [{\citenamefont {Shah}(1996)}]{ShahBook96}%
  \BibitemOpen
  \bibfield  {author} {\bibinfo {author} {\bibfnamefont {J.}~\bibnamefont
  {Shah}},\ }\enquote {\bibinfo {title} {Ultrafast spectroscopy of
  semiconductors and semiconductor nanostructures},}\ \ (\bibinfo  {publisher}
  {Springer},\ \bibinfo {address} {Berlin},\ \bibinfo {year} {1996})\
  Chap.~\bibinfo {chapter} {2}\BibitemShut {NoStop}%
\bibitem [{\citenamefont {Masia}\ \emph {et~al.}(2012)\citenamefont {Masia},
  \citenamefont {Accanto}, \citenamefont {Langbein},\ and\ \citenamefont
  {Borri}}]{MasiaPRL12}%
  \BibitemOpen
  \bibfield  {author} {\bibinfo {author} {\bibfnamefont {F.}~\bibnamefont
  {Masia}}, \bibinfo {author} {\bibfnamefont {N.}~\bibnamefont {Accanto}},
  \bibinfo {author} {\bibfnamefont {W.}~\bibnamefont {Langbein}}, \ and\
  \bibinfo {author} {\bibfnamefont {P.}~\bibnamefont {Borri}},\ }\href@noop {}
  {\bibfield  {journal} {\bibinfo  {journal} {Phys. Rev. Lett.}\ }\textbf
  {\bibinfo {volume} {108}},\ \bibinfo {pages} {087401} (\bibinfo {year}
  {2012})}\BibitemShut {NoStop}%
\bibitem [{\citenamefont {Accanto}\ \emph {et~al.}(2012)\citenamefont
  {Accanto}, \citenamefont {Masia}, \citenamefont {Moreels}, \citenamefont
  {Hens}, \citenamefont {Langbein},\ and\ \citenamefont
  {Borri}}]{AccantoACSN12}%
  \BibitemOpen
  \bibfield  {author} {\bibinfo {author} {\bibfnamefont {N.}~\bibnamefont
  {Accanto}}, \bibinfo {author} {\bibfnamefont {F.}~\bibnamefont {Masia}},
  \bibinfo {author} {\bibfnamefont {I.}~\bibnamefont {Moreels}}, \bibinfo
  {author} {\bibfnamefont {Z.}~\bibnamefont {Hens}}, \bibinfo {author}
  {\bibfnamefont {W.}~\bibnamefont {Langbein}}, \ and\ \bibinfo {author}
  {\bibfnamefont {P.}~\bibnamefont {Borri}},\ }\href {\doibase
  10.1021/nn300992a} {\bibfield  {journal} {\bibinfo  {journal} {ACS Nano}\
  }\textbf {\bibinfo {volume} {6}},\ \bibinfo {pages} {5227} (\bibinfo {year}
  {2012})}\BibitemShut {NoStop}%
\bibitem [{\citenamefont {Naeem}\ \emph {et~al.}(2015)\citenamefont {Naeem},
  \citenamefont {Masia}, \citenamefont {Christodoulou}, \citenamefont
  {Moreels}, \citenamefont {Borri},\ and\ \citenamefont
  {Langbein}}]{NaeemPRB15}%
  \BibitemOpen
  \bibfield  {author} {\bibinfo {author} {\bibfnamefont {A.}~\bibnamefont
  {Naeem}}, \bibinfo {author} {\bibfnamefont {F.}~\bibnamefont {Masia}},
  \bibinfo {author} {\bibfnamefont {S.}~\bibnamefont {Christodoulou}}, \bibinfo
  {author} {\bibfnamefont {I.}~\bibnamefont {Moreels}}, \bibinfo {author}
  {\bibfnamefont {P.}~\bibnamefont {Borri}}, \ and\ \bibinfo {author}
  {\bibfnamefont {W.}~\bibnamefont {Langbein}},\ }\href {\doibase
  10.1103/PhysRevB.91.121302} {\bibfield  {journal} {\bibinfo  {journal} {Phys.
  Rev. B}\ }\textbf {\bibinfo {volume} {91}},\ \bibinfo {pages} {121302}
  (\bibinfo {year} {2015})}\BibitemShut {NoStop}%
\bibitem [{\citenamefont {Arora}\ \emph {et~al.}(2015)\citenamefont {Arora},
  \citenamefont {Nogajewski}, \citenamefont {Molas}, \citenamefont {Koperski},\
  and\ \citenamefont {Potemski}}]{AroraNS15}%
  \BibitemOpen
  \bibfield  {author} {\bibinfo {author} {\bibfnamefont {A.}~\bibnamefont
  {Arora}}, \bibinfo {author} {\bibfnamefont {K.}~\bibnamefont {Nogajewski}},
  \bibinfo {author} {\bibfnamefont {M.}~\bibnamefont {Molas}}, \bibinfo
  {author} {\bibfnamefont {M.}~\bibnamefont {Koperski}}, \ and\ \bibinfo
  {author} {\bibfnamefont {M.}~\bibnamefont {Potemski}},\ }\href {\doibase
  10.1039/C5NR06782K} {\bibfield  {journal} {\bibinfo  {journal} {Nanoscale}\
  }\textbf {\bibinfo {volume} {7}},\ \bibinfo {pages} {20769} (\bibinfo {year}
  {2015})}\BibitemShut {NoStop}%
\bibitem [{\citenamefont {Gartstein}\ \emph {et~al.}(2015)\citenamefont
  {Gartstein}, \citenamefont {Li},\ and\ \citenamefont
  {Zhang}}]{GartsteinPRB15}%
  \BibitemOpen
  \bibfield  {author} {\bibinfo {author} {\bibfnamefont {Y.~N.}\ \bibnamefont
  {Gartstein}}, \bibinfo {author} {\bibfnamefont {X.}~\bibnamefont {Li}}, \
  and\ \bibinfo {author} {\bibfnamefont {C.}~\bibnamefont {Zhang}},\ }\href
  {\doibase 10.1103/PhysRevB.92.075445} {\bibfield  {journal} {\bibinfo
  {journal} {Phys. Rev. B}\ }\textbf {\bibinfo {volume} {92}},\ \bibinfo
  {pages} {075445} (\bibinfo {year} {2015})}\BibitemShut {NoStop}%
\bibitem [{\citenamefont {Wang}\ \emph {et~al.}(2016)\citenamefont {Wang},
  \citenamefont {Zhang}, \citenamefont {Chan}, \citenamefont {Manolatou},
  \citenamefont {Tiwari},\ and\ \citenamefont {Rana}}]{WangPRB16}%
  \BibitemOpen
  \bibfield  {author} {\bibinfo {author} {\bibfnamefont {H.}~\bibnamefont
  {Wang}}, \bibinfo {author} {\bibfnamefont {C.}~\bibnamefont {Zhang}},
  \bibinfo {author} {\bibfnamefont {W.}~\bibnamefont {Chan}}, \bibinfo {author}
  {\bibfnamefont {C.}~\bibnamefont {Manolatou}}, \bibinfo {author}
  {\bibfnamefont {S.}~\bibnamefont {Tiwari}}, \ and\ \bibinfo {author}
  {\bibfnamefont {F.}~\bibnamefont {Rana}},\ }\href@noop {} {\bibfield
  {journal} {\bibinfo  {journal} {Phys. Rev. B}\ }\textbf {\bibinfo {volume}
  {93}},\ \bibinfo {pages} {045407} (\bibinfo {year} {2016})}\BibitemShut
  {NoStop}%
\bibitem [{\citenamefont {Andreani}\ and\ \citenamefont
  {Bassani}(1990)}]{AndreaniPRB90}%
  \BibitemOpen
  \bibfield  {author} {\bibinfo {author} {\bibfnamefont {L.~C.}\ \bibnamefont
  {Andreani}}\ and\ \bibinfo {author} {\bibfnamefont {F.}~\bibnamefont
  {Bassani}},\ }\href@noop {} {\bibfield  {journal} {\bibinfo  {journal} {Phys.
  Rev. B}\ }\textbf {\bibinfo {volume} {41}},\ \bibinfo {pages} {7536}
  (\bibinfo {year} {1990})}\BibitemShut {NoStop}%
\bibitem [{\citenamefont {Andreani}(1995)}]{AndreaniBook94}%
  \BibitemOpen
  \bibfield  {author} {\bibinfo {author} {\bibfnamefont {L.~C.}\ \bibnamefont
  {Andreani}},\ }\enquote {\bibinfo {title} {Confined electrons and photons:
  New physics and applications},}\ \ (\bibinfo  {publisher} {Plenum Press},\
  \bibinfo {address} {New York},\ \bibinfo {year} {1995})\ pp.\ \bibinfo
  {pages} {57--112}\BibitemShut {NoStop}%
\bibitem [{\citenamefont {Dufferwiel}\ \emph {et~al.}(2015)\citenamefont
  {Dufferwiel}, \citenamefont {Schwarz}, \citenamefont {Withers}, \citenamefont
  {Trichet}, \citenamefont {Li}, \citenamefont {Sich}, \citenamefont
  {Pozo-Zamudio}, \citenamefont {Clark}, \citenamefont {Nalitov}, \citenamefont
  {Solnyshkov}, \citenamefont {Malpuech}, \citenamefont {Novoselov},
  \citenamefont {Smith}, \citenamefont {Skolnick}, \citenamefont
  {Krizhanovskii},\ and\ \citenamefont {Tartakovskii}}]{DufferwielNC15}%
  \BibitemOpen
  \bibfield  {author} {\bibinfo {author} {\bibfnamefont {S.}~\bibnamefont
  {Dufferwiel}}, \bibinfo {author} {\bibfnamefont {S.}~\bibnamefont {Schwarz}},
  \bibinfo {author} {\bibfnamefont {F.}~\bibnamefont {Withers}}, \bibinfo
  {author} {\bibfnamefont {A.}~\bibnamefont {Trichet}}, \bibinfo {author}
  {\bibfnamefont {F.}~\bibnamefont {Li}}, \bibinfo {author} {\bibfnamefont
  {M.}~\bibnamefont {Sich}}, \bibinfo {author} {\bibfnamefont {O.~D.}\
  \bibnamefont {Pozo-Zamudio}}, \bibinfo {author} {\bibfnamefont
  {C.}~\bibnamefont {Clark}}, \bibinfo {author} {\bibfnamefont
  {A.}~\bibnamefont {Nalitov}}, \bibinfo {author} {\bibfnamefont
  {D.}~\bibnamefont {Solnyshkov}}, \bibinfo {author} {\bibfnamefont
  {G.}~\bibnamefont {Malpuech}}, \bibinfo {author} {\bibfnamefont
  {K.}~\bibnamefont {Novoselov}}, \bibinfo {author} {\bibfnamefont
  {J.}~\bibnamefont {Smith}}, \bibinfo {author} {\bibfnamefont
  {M.}~\bibnamefont {Skolnick}}, \bibinfo {author} {\bibfnamefont
  {D.}~\bibnamefont {Krizhanovskii}}, \ and\ \bibinfo {author} {\bibfnamefont
  {A.}~\bibnamefont {Tartakovskii}},\ }\href@noop {} {\bibfield  {journal}
  {\bibinfo  {journal} {Nat. Commun.}\ }\textbf {\bibinfo {volume} {6}},\
  \bibinfo {pages} {8579} (\bibinfo {year} {2015})}\BibitemShut {NoStop}%
\bibitem [{\citenamefont {Savona}\ and\ \citenamefont
  {Langbein}(2006)}]{SavonaPRB06}%
  \BibitemOpen
  \bibfield  {author} {\bibinfo {author} {\bibfnamefont {V.}~\bibnamefont
  {Savona}}\ and\ \bibinfo {author} {\bibfnamefont {W.}~\bibnamefont
  {Langbein}},\ }\href@noop {} {\bibfield  {journal} {\bibinfo  {journal}
  {Phys. Rev. B}\ }\textbf {\bibinfo {volume} {74}},\ \bibinfo {pages} {075311}
  (\bibinfo {year} {2006})}\BibitemShut {NoStop}%
\bibitem [{\citenamefont {Zimmermann}\ \emph {et~al.}(2003)\citenamefont
  {Zimmermann}, \citenamefont {Runge},\ and\ \citenamefont
  {Savona}}]{ZimmermannBook03}%
  \BibitemOpen
  \bibfield  {author} {\bibinfo {author} {\bibfnamefont {R.}~\bibnamefont
  {Zimmermann}}, \bibinfo {author} {\bibfnamefont {E.}~\bibnamefont {Runge}}, \
  and\ \bibinfo {author} {\bibfnamefont {V.}~\bibnamefont {Savona}},\ }in\
  \href@noop {} {\emph {\bibinfo {booktitle} {Quantum Coherence, Correlation
  and Decoherence in Semiconductor Nanostructures}}},\ \bibinfo {editor}
  {edited by\ \bibinfo {editor} {\bibfnamefont {T.}~\bibnamefont
  {Takagahara}}}\ (\bibinfo  {publisher} {Elsevier Science},\ \bibinfo
  {address} {USA},\ \bibinfo {year} {2003})\ p.~\bibinfo {pages}
  {89}\BibitemShut {NoStop}%
\bibitem [{\citenamefont {Zimmermann}\ \emph {et~al.}(2001)\citenamefont
  {Zimmermann}, \citenamefont {Langbein}, \citenamefont {Runge},\ and\
  \citenamefont {Hvam}}]{ZimmermannPE01}%
  \BibitemOpen
  \bibfield  {author} {\bibinfo {author} {\bibfnamefont {R.}~\bibnamefont
  {Zimmermann}}, \bibinfo {author} {\bibfnamefont {W.}~\bibnamefont
  {Langbein}}, \bibinfo {author} {\bibfnamefont {E.}~\bibnamefont {Runge}}, \
  and\ \bibinfo {author} {\bibfnamefont {J.~M.}\ \bibnamefont {Hvam}},\
  }\bibfield  {booktitle} {\emph {\bibinfo {booktitle} {Proceedings of the 24th
  {ICPS}}},\ }\href@noop {} {\bibfield  {journal} {\bibinfo  {journal} {Physica
  E}\ }\textbf {\bibinfo {volume} {10}},\ \bibinfo {pages} {40} (\bibinfo
  {year} {2001})}\BibitemShut {NoStop}%
\bibitem [{\citenamefont {Godde}\ \emph {et~al.}(2016)\citenamefont {Godde},
  \citenamefont {Schmidt}, \citenamefont {Schmutzler}, \citenamefont {Abmann},
  \citenamefont {Debus}, \citenamefont {Withers}, \citenamefont {Alexeev},
  \citenamefont {Pozo-Zamudio}, \citenamefont {Skrypka}, \citenamefont
  {Novoselov}, \citenamefont {Bayer},\ and\ \citenamefont
  {Tartakovskii}}]{GoddePRB16}%
  \BibitemOpen
  \bibfield  {author} {\bibinfo {author} {\bibfnamefont {T.}~\bibnamefont
  {Godde}}, \bibinfo {author} {\bibfnamefont {D.}~\bibnamefont {Schmidt}},
  \bibinfo {author} {\bibfnamefont {J.}~\bibnamefont {Schmutzler}}, \bibinfo
  {author} {\bibfnamefont {M.}~\bibnamefont {Abmann}}, \bibinfo {author}
  {\bibfnamefont {J.}~\bibnamefont {Debus}}, \bibinfo {author} {\bibfnamefont
  {F.}~\bibnamefont {Withers}}, \bibinfo {author} {\bibfnamefont
  {E.}~\bibnamefont {Alexeev}}, \bibinfo {author} {\bibfnamefont {O.~D.}\
  \bibnamefont {Pozo-Zamudio}}, \bibinfo {author} {\bibfnamefont
  {O.}~\bibnamefont {Skrypka}}, \bibinfo {author} {\bibfnamefont
  {K.}~\bibnamefont {Novoselov}}, \bibinfo {author} {\bibfnamefont
  {M.}~\bibnamefont {Bayer}}, \ and\ \bibinfo {author} {\bibfnamefont
  {A.}~\bibnamefont {Tartakovskii}},\ }\href@noop {} {\bibfield  {journal}
  {\bibinfo  {journal} {Phys. Rev. B}\ }\textbf {\bibinfo {volume} {94}},\
  \bibinfo {pages} {165301} (\bibinfo {year} {2016})}\BibitemShut {NoStop}%
\bibitem [{\citenamefont {Yu}\ \emph {et~al.}(2016)\citenamefont {Yu},
  \citenamefont {Yu}, \citenamefont {Xu}, \citenamefont {Barrette},
  \citenamefont {Gundogdu},\ and\ \citenamefont {Cao}}]{YuPRB16}%
  \BibitemOpen
  \bibfield  {author} {\bibinfo {author} {\bibfnamefont {Y.}~\bibnamefont
  {Yu}}, \bibinfo {author} {\bibfnamefont {Y.}~\bibnamefont {Yu}}, \bibinfo
  {author} {\bibfnamefont {C.}~\bibnamefont {Xu}}, \bibinfo {author}
  {\bibfnamefont {A.}~\bibnamefont {Barrette}}, \bibinfo {author}
  {\bibfnamefont {K.}~\bibnamefont {Gundogdu}}, \ and\ \bibinfo {author}
  {\bibfnamefont {L.}~\bibnamefont {Cao}},\ }\href@noop {} {\bibfield
  {journal} {\bibinfo  {journal} {Phys. Rev. B}\ }\textbf {\bibinfo {volume}
  {96}},\ \bibinfo {pages} {201111} (\bibinfo {year} {2016})}\BibitemShut
  {NoStop}%
\bibitem [{\citenamefont {Shi}\ \emph {et~al.}(2013)\citenamefont {Shi},
  \citenamefont {Yan}, \citenamefont {Bertolazzi}, \citenamefont {Brivio},
  \citenamefont {Gao}, \citenamefont {Kis}, \citenamefont {Jena}, \citenamefont
  {Xing},\ and\ \citenamefont {Huang}}]{ShiACSN13}%
  \BibitemOpen
  \bibfield  {author} {\bibinfo {author} {\bibfnamefont {H.}~\bibnamefont
  {Shi}}, \bibinfo {author} {\bibfnamefont {R.}~\bibnamefont {Yan}}, \bibinfo
  {author} {\bibfnamefont {S.}~\bibnamefont {Bertolazzi}}, \bibinfo {author}
  {\bibfnamefont {J.}~\bibnamefont {Brivio}}, \bibinfo {author} {\bibfnamefont
  {B.}~\bibnamefont {Gao}}, \bibinfo {author} {\bibfnamefont {A.}~\bibnamefont
  {Kis}}, \bibinfo {author} {\bibfnamefont {D.}~\bibnamefont {Jena}}, \bibinfo
  {author} {\bibfnamefont {H.~G.}\ \bibnamefont {Xing}}, \ and\ \bibinfo
  {author} {\bibfnamefont {L.}~\bibnamefont {Huang}},\ }\href@noop {}
  {\bibfield  {journal} {\bibinfo  {journal} {ACS Nano}\ }\textbf {\bibinfo
  {volume} {2}},\ \bibinfo {pages} {1072} (\bibinfo {year} {2013})}\BibitemShut
  {NoStop}%
\bibitem [{\citenamefont {Wang}\ \emph
  {et~al.}(2015{\natexlab{b}})\citenamefont {Wang}, \citenamefont {Zhang},\
  and\ \citenamefont {Rana}}]{WangNL15}%
  \BibitemOpen
  \bibfield  {author} {\bibinfo {author} {\bibfnamefont {H.}~\bibnamefont
  {Wang}}, \bibinfo {author} {\bibfnamefont {C.}~\bibnamefont {Zhang}}, \ and\
  \bibinfo {author} {\bibfnamefont {F.}~\bibnamefont {Rana}},\ }\href@noop {}
  {\bibfield  {journal} {\bibinfo  {journal} {Nano Lett.}\ }\textbf {\bibinfo
  {volume} {15}},\ \bibinfo {pages} {339} (\bibinfo {year}
  {2015}{\natexlab{b}})}\BibitemShut {NoStop}%
\bibitem [{Kum()}]{KumarPRB14Note}%
  \BibitemOpen
  \href@noop {} {}\bibinfo {note} {In the solution given in Ref. [9] equation
  (3), a factor 1/2 is missing}\BibitemShut {NoStop}%
\bibitem [{\citenamefont {Singh}\ \emph {et~al.}(2016)\citenamefont {Singh},
  \citenamefont {Moody}, \citenamefont {Tran}, \citenamefont {Scott},
  \citenamefont {Overbeck}, \citenamefont {Bergh\"{a}user}, \citenamefont
  {Schaibley}, \citenamefont {Seifert}, \citenamefont {Pleskot}, \citenamefont
  {Gabor}, \citenamefont {Yan}, \citenamefont {Mandrus}, \citenamefont
  {Richter}, \citenamefont {Malic}, \citenamefont {Xu},\ and\ \citenamefont
  {Li1}}]{SinghPRB16}%
  \BibitemOpen
  \bibfield  {author} {\bibinfo {author} {\bibfnamefont {A.}~\bibnamefont
  {Singh}}, \bibinfo {author} {\bibfnamefont {G.}~\bibnamefont {Moody}},
  \bibinfo {author} {\bibfnamefont {K.}~\bibnamefont {Tran}}, \bibinfo {author}
  {\bibfnamefont {M.~E.}\ \bibnamefont {Scott}}, \bibinfo {author}
  {\bibfnamefont {V.}~\bibnamefont {Overbeck}}, \bibinfo {author}
  {\bibfnamefont {G.}~\bibnamefont {Bergh\"{a}user}}, \bibinfo {author}
  {\bibfnamefont {J.}~\bibnamefont {Schaibley}}, \bibinfo {author}
  {\bibfnamefont {E.~J.}\ \bibnamefont {Seifert}}, \bibinfo {author}
  {\bibfnamefont {D.}~\bibnamefont {Pleskot}}, \bibinfo {author} {\bibfnamefont
  {N.~M.}\ \bibnamefont {Gabor}}, \bibinfo {author} {\bibfnamefont
  {J.}~\bibnamefont {Yan}}, \bibinfo {author} {\bibfnamefont {D.~G.}\
  \bibnamefont {Mandrus}}, \bibinfo {author} {\bibfnamefont {M.}~\bibnamefont
  {Richter}}, \bibinfo {author} {\bibfnamefont {E.}~\bibnamefont {Malic}},
  \bibinfo {author} {\bibfnamefont {X.}~\bibnamefont {Xu}}, \ and\ \bibinfo
  {author} {\bibfnamefont {X.}~\bibnamefont {Li1}},\ }\href@noop {} {\bibfield
  {journal} {\bibinfo  {journal} {Phys. Rev. B}\ }\textbf {\bibinfo {volume}
  {93}},\ \bibinfo {pages} {041401(R)} (\bibinfo {year} {2016})}\BibitemShut
  {NoStop}%
\bibitem [{\citenamefont {Jones}\ \emph {et~al.}(2016)\citenamefont {Jones},
  \citenamefont {Yu}, \citenamefont {Schaibley}, \citenamefont {Yan},
  \citenamefont {Mandrus}, \citenamefont {Taniguchi}, \citenamefont {Watanabe},
  \citenamefont {Dery}, \citenamefont {Yao},\ and\ \citenamefont
  {Xu}}]{JonesNPhy15}%
  \BibitemOpen
  \bibfield  {author} {\bibinfo {author} {\bibfnamefont {A.~M.}\ \bibnamefont
  {Jones}}, \bibinfo {author} {\bibfnamefont {H.}~\bibnamefont {Yu}}, \bibinfo
  {author} {\bibfnamefont {J.~R.}\ \bibnamefont {Schaibley}}, \bibinfo {author}
  {\bibfnamefont {J.}~\bibnamefont {Yan}}, \bibinfo {author} {\bibfnamefont
  {D.~G.}\ \bibnamefont {Mandrus}}, \bibinfo {author} {\bibfnamefont
  {T.}~\bibnamefont {Taniguchi}}, \bibinfo {author} {\bibfnamefont
  {K.}~\bibnamefont {Watanabe}}, \bibinfo {author} {\bibfnamefont
  {H.}~\bibnamefont {Dery}}, \bibinfo {author} {\bibfnamefont {W.}~\bibnamefont
  {Yao}}, \ and\ \bibinfo {author} {\bibfnamefont {X.}~\bibnamefont {Xu}},\
  }\href {\doibase 10.1038/NPHYS3604} {\bibfield  {journal} {\bibinfo
  {journal} {Nat. Phys.}\ }\textbf {\bibinfo {volume} {12}},\ \bibinfo {pages}
  {323} (\bibinfo {year} {2016})}\BibitemShut {NoStop}%
\end{thebibliography}%

\end{document}